\documentclass{aa}  
\usepackage{amsmath}
\usepackage{natbib}
\usepackage{units}
\usepackage{color}

\usepackage{graphicx}
\usepackage{txfonts}

\newcommand{\pd}[2]{\frac{\partial #1}{\partial #2} }

\newcommand{\beq}{\begin{equation}}
\newcommand{\eeq}{\end{equation}}
\newcommand{\AU}{\text{A\!U}}

\begin{document}
	\title{Vertical shear instability in accretion disc models with radiation transport}
	\author{Moritz H. R. Stoll
	\and
        Wilhelm Kley
	}

	\institute{
	Institut f\"ur Astronomie und Astrophysik, Universit\"at 
	T\"ubingen, Auf der Morgenstelle 10, D-72076 T\"ubingen, Germany\\
	\email{moritz.stoll@student.uni-tuebingen.de,wilhelm.kley@uni-tuebingen.de}\\
	}

 	\date{Received 2014-05-05; accepted 2014-09-25}

\abstract
        {The origin of turbulence in accretion discs is still not fully understood. While the magneto-rotational
        instability is considered to operate in sufficiently ionized discs, its role in the poorly ionized
        protoplanetary disc is questionable. Recently, the vertical shear instability (VSI) has been
        suggested as a possible alternative. 
        }
        {Our goal is to study the characteristics of this instability and the efficiency of angular momentum transport,
         in extended discs, under the influence of radiative transport and 
         irradiation from the central star.
 	}
 	{We use multi-dimensional hydrodynamic simulations to model a larger section of an accretion disc.
         First we study inviscid and weakly viscous discs using a fixed radial temperature profile in two and three
         spatial dimensions. The simulations are then extended to include radiative transport and irradiation from the
         central star. 
        }
	{In agreement with previous studies we find for the isothermal disc a sustained unstable state with a weak positive
         angular momentum transport of the order of $\alpha \approx 10^{-4}$. Under the inclusion of radiative transport the disc
         cools off and the turbulence terminates. 
          For discs irradiated from the central star we find again a persistent instability with a similar $\alpha$ value
          as for the isothermal case.
        }
        {We find that the VSI can indeed generate sustained turbulence in discs albeit at a relatively low level with
        $\alpha$ about few times $10^{-4}$.
        }
   \maketitle
%

\section{Introduction}
The origin of the angular momentum transport in accretion discs is still not fully understood.
Observationally it is well confirmed that the molecular viscosity is by many orders of magnitude too small
to explain the effective mass and angular momentum transport in discs \citep{1981ARA&A..19..137P}. This can be inferred for example from time variations
of the disc luminosity in close binary systems, or by correlating the infrared-excess caused by discs around young stars with the age of the system.
As a consequence it is assumed that discs are driven by some kind of turbulent transport whose cause is still not known. 
Despite its unknown origin the efficiency of the turbulence is usually parameterized in terms of the dimensionless parameter, $\alpha$,
as introduced by \citet{Shakura1973}.
Observationally, values of a few times $10^{-3}$ as in protostellar discs to $10^{-1}$ for discs in close binary stars are suggested.
For sufficiently well ionized
discs the magnetorotational instability (MRI) is certainly the most promising candidate to provide the transport \citep{2003ARA&A..41..555B}.
While this may be true for the hot discs in close binary systems or in active galactic nuclei, there is the important class of protostellar discs
where at least the thermal ionisation levels are too low to provide a sufficient number of charged particles that can support the MRI \citep{2011ARA&A..49..195A}.
In such discs the turbulence plays an important role in several aspects. Not only does it determine the lifetime of an accretion disc, but it also influences where and how planets can form and evolve in the disc. 
A variety of sources such as stellar X-rays, cosmic rays or collisions with beta particles from radioactive nuclei have been invoked to provide the required ionization levels
but recent studies indicate the presence of an extended 'dead zone' where, due to the lack of ionization, no magnetically driven instability 
may operate.
Additionally, recent studies on the origins of turbulence in protostellar discs that include non-ideal magnetohydrodynamical (MHD) effects such as ambipolar 
diffusion or the Hall effect, indicate that the MRI may even be suppressed strongly in these discs, see the review by \citet{2014arXiv1401.7306T} and references
therein.

As a consequence alternative mechanisms to provide turbulence are actively discussed.
Typical examples for non-magnetized discs are the convective instability \citep{1988ApJ...329..739R}, the gravitational instability \citep{1987MNRAS.225..607L},
or the baroclinic instability \citep{2003ApJ...582..869K}, for further references see \citet{2013MNRAS.435.2610N}.
While any of these may operate under special conditions in the disc, e.g. suitable radial entropy gradients or a sufficiently high
disc mass, none seems to have general applicability.
Searching for alternatives the vertical shear instability (VSI) has attracted recent interest. Here, the instability is caused by a vertical gradient of the
angular velocity, $\Omega$, in the disc. Through linear analysis it has been shown that for a sufficiently strong vertical shear there are always modes that 
can overcome the stabilizing angular momentum gradient (Rayleigh-criterion) and generate instability \citep{1998MNRAS.294..399U, 2003A&A...404..397U}. 
This instability is related to the Goldreich-Schubert-Fricke instability that can occur in differentially rotating stars \citep{1967ApJ...150..571G, 1968ZA.....68..317F}.  

Concerning its effectiveness with respect to angular momentum transport numerical simulations were performed by \citet{2004A&A...426..755A} and \citet{2013MNRAS.435.2610N}.
The first authors analysed the instability for globally isothermal discs and found that the instability in this case could only be triggered by applying
finite initial perturbation because the equilibrium state of the disc (being strictly isothermal) did not contain a shear in $\Omega$.
The maximum values of $\alpha$ obtained by \citet{2004A&A...426..755A} were around $6 \cdot 10^{-6}$ but the turbulence was decaying in the long run. 
\citet{2013MNRAS.435.2610N} extended these simulations and performed high resolution simulations of the VSI for so called locally isothermal discs
that contain a radial temperature gradient but are vertically isothermal. Under these conditions the equilibrium state has a vertical gradient in the shear and
indeed an instability sets in. As shown by \citet{2013MNRAS.435.2610N} the instability has two distinct growth phases, 
it starts from the surface layers of the disc where the shear is strongest and then protrudes towards the midplane.
In the final state the vertical motions in the disc are antisymmetric with respect to the discs's midplane, such that the gas elements cross
the midplane, a feature found for the vertical convective motions in discs as well \citep{1993ApJ...416..679K}.
For the efficiency of the VSI induced turbulence \citet{2013MNRAS.435.2610N} found a weak angular momentum transport with $\alpha =6\cdot 10^{-4}$. 
They also showed that under the presence of a small viscosity or thermal relaxation the instability is weaker and can easily be quenched.
   
It is not clear what influence radiation transport will have on this instability. 
Without external heat sources one might expect that, because of radiative cooling and the dependence of the instability on temperature, the instability will die out. Here, we evaluate the evolution of the instability for a radiative discs and an ideal equation of state. 
Additionally, we extend the radial domain and include irradiation from the central star.
We perform two-dimensional and three-dimensional hydrodynamical simulations including radiative transport.

This paper is organized as follows. In Section 2, we present the physical setup of our disc models and in Section 3 the numerical approach.
The isothermal results are presented in Section 4, followed by the radiative
cases in Section 5. Stellar irradiation is considered in Section 6
and in Section 7 we conclude.

\section{Physical setup}
In order to study the VSI of the disc in the presence of radiative transport we 
construct numerical models solving the hydrodynamical equations for a section of the
accretion disc in two and three spatial dimensions.

\subsection{Equations}
The basis our studies are the Euler equations (\ref{eq:euler1} - \ref{eq:euler3}) describing the motion of an ideal gas.
These are coupled to radiation transport (\ref{eq:rad}) for which we use the two temperature approximation applying
flux-limited diffusion.
The equations then read 
\begin{eqnarray}
\frac{\partial \rho}{\partial t} + \nabla (\rho \vec{u}) &=& 0 \label{eq:euler1} \\
\pd{}{t}  \rho \vec{u} + \nabla (\rho \vec{u}  \vec{u}) + \nabla p &=& \rho \vec{a}_{ext} \label{eq:euler2} \\
\pd{}{t} e + \nabla [(e + p) \vec{u}] &=& \rho \vec{u} \vec{a}_{ext} - \kappa_P \rho c (a_R T^4 -E)\label{eq:euler3}  \\
\pd{}{t} E + \nabla \vec{F} &=& \kappa_P \rho c\left( a_R T^4 -E \right) \label{eq:rad} \,. 
\end{eqnarray}
Here $\rho$ is the density, $\vec{u}$ the velocity, $e$ the total energy density (kinetic and thermal) of the gas. $p$ denotes the gas pressure,
and the acceleration due to external forces, such as the gravitational force exerted by the central star is given by $\vec{a}_{ext}$.
$E$ and $\vec{F}$ are the energy density and the flux of the radiation. The last terms on the r.h.s. of eqs.~(\ref{eq:euler3}) and (\ref{eq:rad}) 
refer to the coupling of gas and radiation, i.e. the heating/cooling terms. Here,
$c$ stands for the speed of light, $a_R$ is the radiation constant, and $\kappa_P$ the Plank mean opacity.

We close the equations with the ideal gas equation of state
\begin{equation}
\label{eq:eos}
   p = (\gamma-1) e_{th} \, ,
\end{equation}
where $e_{th} = e - 1/2 \rho u^2$ is the thermal energy density.
The temperature of the gas is then calculated from
\begin{equation}
\label{eq:eostemp}
   p =\rho \frac{k_B T}{\mu m_H} \, ,
\end{equation}
where  $\mu$ is the mean molecular weight, $k_B$ the Boltzmann constant and $m_H$ the mass of the hydrogen atom.
In our simulations with radiation transport we use $\gamma = 1.4$ and $\mu = 2.35$.
To compare to previous studies  we performed additional isothermal simulations where we use
$\gamma = 1.001$ and additionally reset to the original temperature profile in every step. 
This procedure corresponds to an isothermal simulation but allows for an arbitrary temperature profile. 
It also allows to use the feature of slowly relaxing to a given original temperature such as 
used for example in \citet{2013MNRAS.435.2610N}. Note that without resetting the temperature the gas remains adiabatic, and
the perturbation will die out for our setup.

The radiation flux in the flux-limited diffusion (FLD) approximation \citep{1981ApJ...248..321L}  is given by
\begin{equation}
\vec{F} = - \lambda \frac{c}{\kappa_R \rho} \nabla E \,,
  \label{eq:fld}
\end{equation}
where $\kappa_R$ is the Rosseland mean opacity and $\lambda$ is the flux-limiter, for which we use the description of
\citet{1978JQSRT..20..541M}. For the Rosseland mean opacity we apply the model of \citet{Bell1994}. For simplicity we use in this initial study
the same value for the Plank mean opacity, see also \citet{2013A&A...549A.124B}.

In some of our studies we add viscosity and stellar irradiation to the momentum and energy equations. This will be pointed out below in the appropriate sections.

\subsection{Disc model}
To be able to study the onset of the instability we start with a reference model in equilibrium. For this purpose,
we follow \citet{2013MNRAS.435.2610N} and use a locally isothermal disc in force equilibrium, where 
for the midplane density we assume a power law behaviour
\begin{equation}
  \rho(R,Z=0) = \rho_0 \left( \frac{R}{R_0} \right)^p  \,,
\label{eq:rho0}
\end{equation}
and the temperature is constant on cylinders 
\begin{equation}
  T(R,Z) = T_0 \left( \frac{R}{R_0} \right)^q \,.
\label{eq:Temp0}
\end{equation}
To specify the equilibrium state we have used a cylindrical coordinate system ($R,Z,\phi$). However, our simulations will be
performed in spherical polar coordinates ($r,\theta,\phi$) because they are better adapted to the geometry of an accretion disc. 
In eqs.~(\ref{eq:rho0}) and (\ref{eq:Temp0}), $\rho_0$ and $T_0$ are suitably
chosen constants that determine the total mass content in the disc and its temperature. The exponents $p$ and $q$ give the radial
steepness of the profiles, and typically we choose $p=-3/2$ and $q=-1$.  
Assuming that in the initial state there are no motions in the meridional plane and the flow is purely toroidal, 
force balance in the radial and vertical directions then leads to the equilibrium density and angular velocity profiles that we use for the initial setup
\citep{2013MNRAS.435.2610N} 
\begin{equation}
  \rho(R,Z) = \rho_0 \left( \frac{R}{R_0} \right)^p \,  \exp{ \left[ \frac{G M}{c_s^2}\left( \frac{1}{\sqrt{R^2 + Z^2}} - \frac{1}{R} \right) \right]} \,,
\label{eq:rho2d}
\end{equation}
and
\begin{equation}
\Omega(R,Z) = \Omega_K \left[ (p +q)\left( \frac{H}{R} \right)^2 + (1+q) - \frac{qR}{\sqrt{R^2 + Z^2}} \right]^{\frac{1}{2}} \,.
\label{equ:omega}
\end{equation}
Here, $c_s = \sqrt{p/\rho}$ denotes the isothermal sound speed, $\Omega_K =
\sqrt{G M_{\odot}/R^3}$ the Keplerian angular velocity, and
$H = c_s / \Omega_K$ is the local pressure scale height of the accretion disc. 
We note that the $Z$ dependence of $\Omega$ in the equilibrium state is the origin of the VSI, because the vertical shear provides the opportunity for fluid perturbations with a wavenumber ratio $k_R/k_Z$ above a threshold to tap into a negative gradient in the angular momentum as the perturbed fluid elements move away from the rotation axis \citep{2013MNRAS.435.2610N}.
The angular velocity given by eq.~(\ref{equ:omega}) is also used to calculate the Reynolds stress tensor, for details see below.

\subsection{Stability}
\citet{2013MNRAS.435.2610N} repeated the original analysis in \citet{1967ApJ...150..571G} for a locally isothermal and compressive gas for an
accretion disc using the local shearing sheet approximation at a reference radius $r_0$. They derived the
same stability criterion as \citet{2003A&A...404..397U} and obtained the following growth rate of the instability 
\begin{equation}
  \sigma^2 = \frac{-\kappa_0^2 (c_0^2 k_Z^2 + N_0^2) + 2 \Omega_0 c_0^2 k_R k_Z \frac{\partial \bar{V}}{\partial z}}{c_0^2 (k_Z^2 + k_R^2) + \kappa_0^2 + N_0^2} \,,
\label{eq:sigma0}
\end{equation}
where $\kappa_0$ is the epicyclic frequency, $c_0$ the sound speed, and $N_0$ is the Brunt-Vaisaila frequency at the radius $r_0$.
$\bar{V}$ denotes the mean deviation from of the Keplerian azimuthal velocity profile, and $k_R$, $k_Z$ are the radial and vertical wavenumbers
of the perturbations in the local coordinates. 

For negligible $N_0$, small $H_0/R_0$, and $k_Z/k_R \sim {\cal{O}} (qH_0/R_0)$, as seen in their numerical simulations, \citet{2013MNRAS.435.2610N} find
\begin{equation}
  \sigma \sim  q \Omega \frac{H}{R} \,,
\label{eq:sigma}
\end{equation}
which implies that the growth rate per local orbit to first order depends on the temperature gradient as given by $q$ 
and on the absolute temperature, due to ${H}/{R}$. We will compare our numerical results with these estimates.

\section{Numerical Model}
To study the VSI in the presence of radiative transport we perform numerical simulations of a section of an accretion disc in
two and three spatial dimensions using spherical polar coordinates $(r, \theta, \phi)$, and a grid which is logarithmic in radial direction, keeping the cells squared. 
We solve eqs.~(\ref{eq:euler1}) to (\ref{eq:rad}) with a grid based method, where we 
use the \texttt{PLUTO} code from \citet{2007ApJS..170..228M} that utilizes a second-order Godunov scheme,
together with our radiation transport \citep{2013A&A...559A..80K} in the FLD approximation, see eq.~(\ref{eq:fld}). 

The simulations span a region in radius from $r = 2-10\AU$, this is the range where the dead zone can be expected \citep{2011ARA&A..49..195A,2012MNRAS.420.2419F}.
Here, we use a larger radial domain as \citet{2013MNRAS.435.2610N} because we 
intend to study the global properties of the instability over a wider range of distances. Additionally, this larger range is useful, because we need some 
additional space (typically $\approx 1$\AU) to damp possible large scale vortices in the meridional plane that show up at the inner radial boundary of the domain (see below).
The origin of these vortices is possibly that the instability moves material along cylindrically shaped shells,
a motion that is not adapted to the used spherical coordinates, such that the midplane is cut out at the inner boundary. Vortices
can also arise if the viscosity changes apruptly, a situation mimicking a boundary.
Additionally, in some cases the wavelengths are large, such that the coupling between different modes cannot be captured in a small domain.
Also we use a wide range because with radiation transport the growth rates are expected to depend on the opacity, which is a function of $\rho$ and $T$ and thus of the radius. 
In the meridional direction ($\theta$) we go up to $\pm 5$ scale heights above and below the equator in the isothermal case,
and we use the same extension for the radiative simulation, where it corresponds to more scale heights.
For the 3D simulations we used in the azimuthal direction ($\phi$) a quarter circle, from 0 to $\pi/2$.

We use reflective boundaries in the radial direction. In the meridional direction we use outflow conditions for the flow out of the domain and reflective conditions
otherwise. For the radiation transport solver we set the temperature of the meridional boundary to 10K, which allows the radiation to escape freely. We use damping of the velocity near the inner radial boundary within $2-3$\AU\ to prevent the creation of strong vortices arising through the interaction with the reflecting boundary, which can destroy the simulation.  This is done by adding a small viscosity of $\nu = 2 \cdot 10^{-7}$ with a linear decrease to zero from 2\AU\ to $3\AU$ (similar to the damping used in \citet{2006MNRAS.370..529D}). 

We assume that the disc orbits a solar mass star and  
we apply a density of $\rho_0 = 10^{-10}${g}/{cm$^3$} at $1\AU$. Due to the
surface density decaying with $r^{-0.5}$ we get a surface density $\Sigma =
80${g}/{cm$^2$} at 5\AU. To study the mass dependence we vary $\rho_0$ for the radiative models.
To seed the instability we add a small perturbation of up to 1\% of the sound speed to the equilibrium velocity, see eq.~(\ref{equ:omega}).

Because our radiation transport solver is only implemented in full 3D \citep{2013A&A...559A..80K}, we use 2 grid cells in the azimuthal direction for the
2D axisymmetric simulations using radiation transport.

\section{Isothermal discs}

\begin{figure}[t]
\begin{center}
\includegraphics{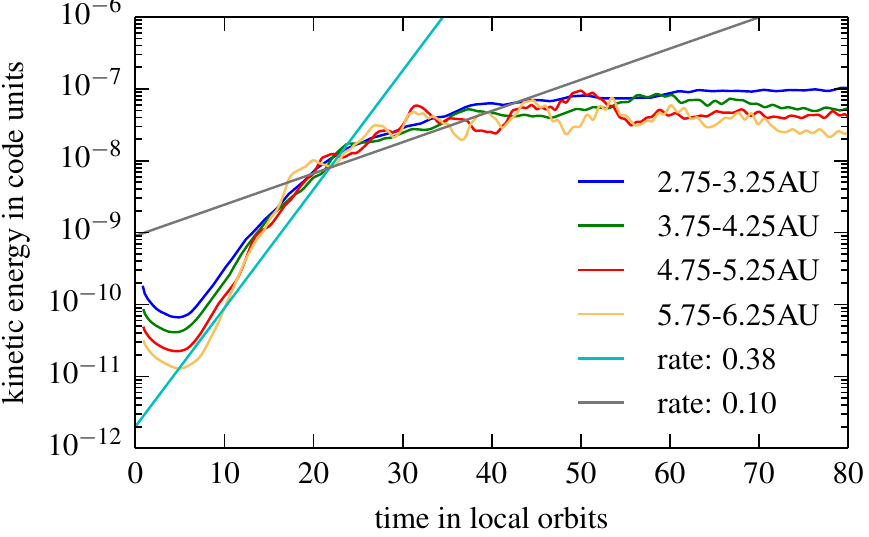}
\end{center}
\caption{The kinetic energy of the motion in the meridional plane at different
radii in an inviscid disc. The kinetic energy at the different locations
is in each case averaged over a radial interval with length 0.5\AU. We note that the unit of time is given in local periods at the center of
the specified interval. Hence, it is different for each curve but this allows for easy comparison.}
\label{fig:growth}
\end{figure}

Before studying full radiative discs we first perform isothermal 2D simulations to compare our results and growth rates to those of \citet{2013MNRAS.435.2610N}. 
Then we will extend the simulation to full 3D using a quarter of a disc and discuss the dependence on resolution and viscosity.

\subsection{Growth rates}
To analyse the possible growth and instability of the initial equilibrium state, we analyse the time evolution of the kinetic energy
in the meridional plane
\begin{equation}
   e_{kin} = \frac{1}{2} \, \rho ( u_r^2 + u^2_\theta ) \,, 
\label{eq:ekin}
\end{equation}
at different radii.
The obtained growth of $e_{kin}$ of a run with $q=-1$ and $p=-3/2$ is displayed at
different radii in Fig.~\ref{fig:growth} for an inviscid disc model with a grid resolution
of $2048 \times 512$. Note that the time is measured in
local orbits ($2 \pi / \Omega(r_i$)) at the corresponding centers of the intervals, $r_i$.  We
measure a mean growth rate of $0.38$ per orbit for the kinetic energy (light
blue line in Fig.~\ref{fig:growth}), which is twice the growth rate ($\sigma$)
of the velocity.  We calculate the growth rate by averaging the kinetic energy
at the different $r_i$ over an interval with length $0.5\AU$. Our results
compare favourably with the growth rates from \citet{2013MNRAS.435.2610N} who obtained
$0.25$ per orbit averaged over $1-2\AU$ for $q=-1$. Averaging over this larger
range leads to a reduced growth because the rate at $2\AU$, measured in orbits
at $1\AU$, is smaller by a factor of $2^{1.5} = 2.8$, and so their result is a
slight underestimate.

\begin{figure}[t]
\begin{center}
\includegraphics{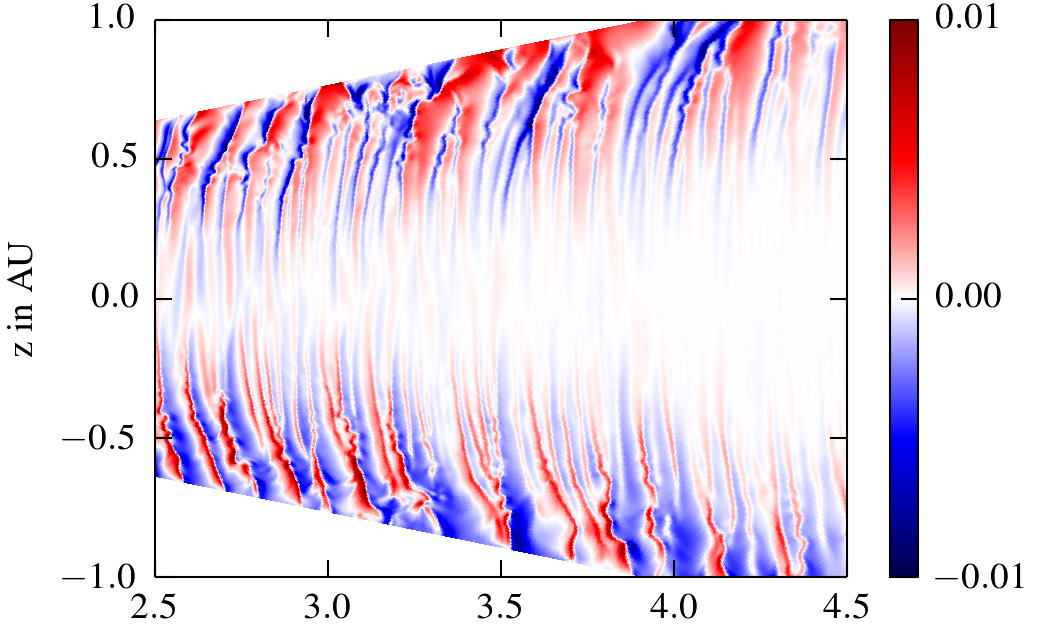}
\includegraphics{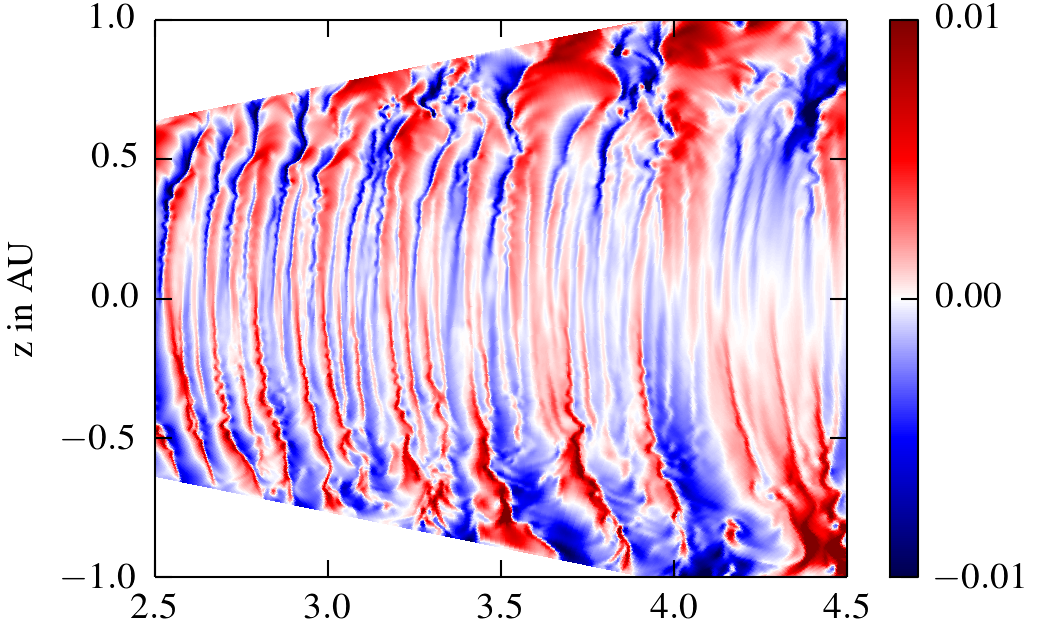}
\includegraphics{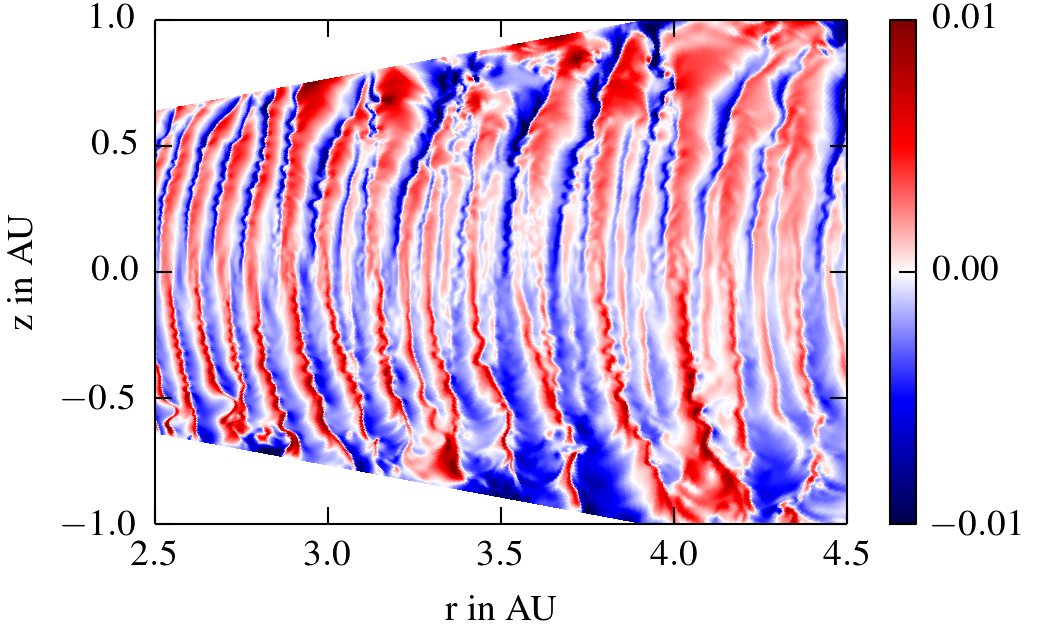}
\end{center}
\caption{Velocity in the meridional direction, $u_\theta$, in units of local Kepler velocity for an isothermal run without viscosity. 
The panels refer to snapshots taken at time 100, 210 and 750 (top to bottom), measured in orbital periods at $1\AU$. In units of local orbits at (2.5, 3.5, 4.5)$\AU$
this refers to (25, 15, 10) (53, 32, 22) (190, 115, 79) orbits, from top to bottom.}
\label{fig:phase1}
\end{figure}

A closer look at figure \ref{fig:growth} reveals two distinct growth phases. An
initial strong linear growth phase with a rate of 0.38 per orbit lasting about 20 local
orbits, and a slower second phase with a rate of 0.10 per orbit (gray line in
Fig.~\ref{fig:growth}). To understand these regimes, we present in
Fig.~\ref{fig:phase1} the velocity in the meridional direction, $u_\theta$, in 2D
contour plots at different times.  The top panel reveals that the first phase
corresponds to symmetric (mirror symmetry with respect to the equatorial
plane) disturbances that grow from the top and bottom surface layers of the
disc. Here, the gas does not cross the midplane of the disc. When those meet
in the disc's midplane they develop an anti-symmetric phase with lower growth rates where the
gas flow crosses the midplane of the disc as displayed in the middle panel.
The converged phase shown in the lower panel then
shows the fully saturated global flow. Figure~\ref{fig:phase1} indicates that
in the top panel the whole domain is still in the anti-symmetric growth phase,
in the middle panel only the smaller radii show symmetric growth, while in the
lower panel the whole domain has reached the final equilibrium, in accordance
with Fig.~\ref{fig:growth}.

We point out that the growth rate per local orbit ($\sim \sigma/\Omega$) is independent of radius in good agreement with the relation (\ref{eq:sigma}), for
constant $H/R$.
We will show later that the growth rate is also independent of resolution.

\subsection{Comparison to 3D results and Reynolds stress}

\begin{figure}[t]
\begin{center}
\includegraphics{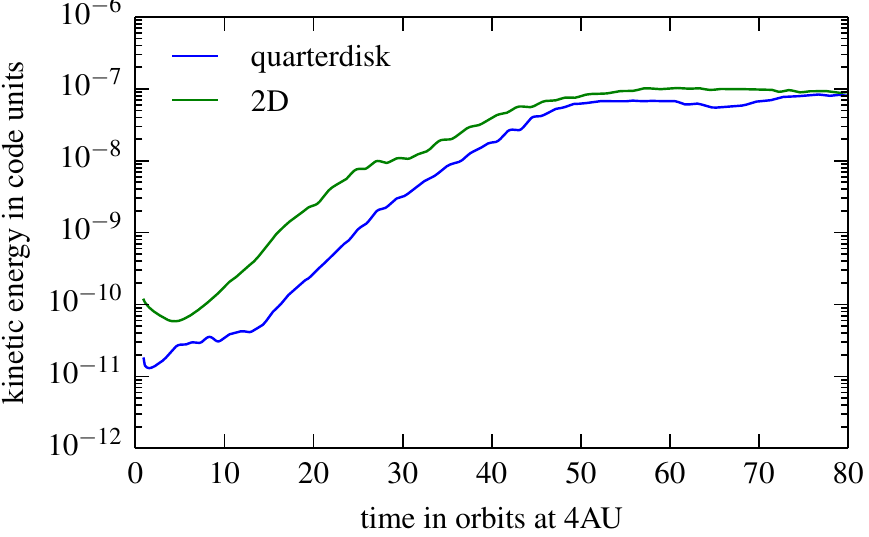}
\end{center}
\caption{The growth of the kinetic energy for the quarter of a disc and the 2D equivalent. The kinetic energy is averaged from 4\AU \ to 5.5\AU.
}
\label{fig:growthdisk}
\end{figure}

In addition to the 2D simulation we ran an equivalent 3D case using a quarter of a disc with a resolution of $512 \times 128 \times 128$ grid cells.  We will use this to discuss the validity of the 2D results, in particular the estimates on the turbulent efficiency factor $\alpha$.  In Fig.~\ref{fig:growthdisk} we compare the growth of the meridional kinetic energy for the 3D and the 2D simulation.  After a slower start, the 3D simulation shows very similar growth and reaches the same final saturation level.

\begin{figure}[tb]
\begin{center}
\includegraphics{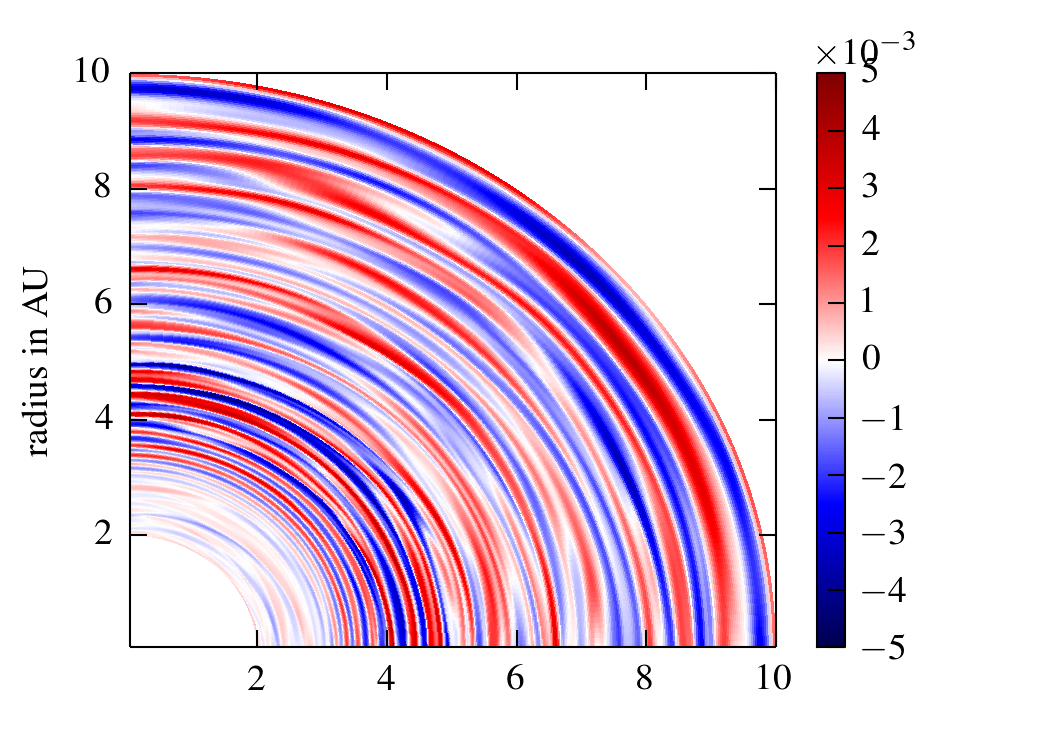}
\end{center}
\caption{The vertical velocity in the midplane of the disk for the 3D model after 4000 orbits. The nearly axisymmetric property of the instability is clearly visible.}
\label{fig:3dmidplane}
\end{figure}

\begin{figure}[tb]
\begin{center}
\includegraphics{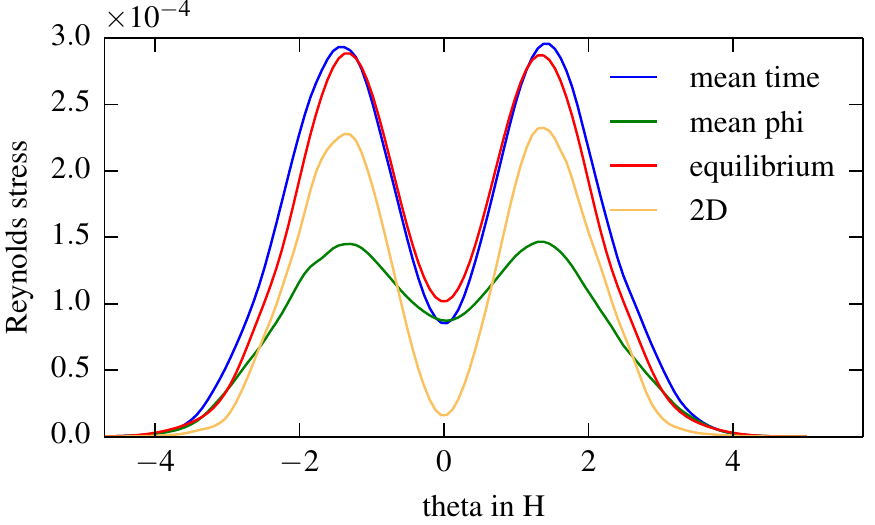}
\end{center}
\caption{The Reynolds stress (code units) from $3\text{-}10\AU$  averaged over 41 time steps, each step 100 orbits apart beginning with orbit 1000, calculated with different averaging methods. 
For 'mean time' the steady state $\bar{u}_{\phi} = u_{\phi} - \delta u_{\phi}$, needed to calculate the Reynolds stress at each step, was calculated through averaging over the 41 time steps. 
For 'mean phi' the steady state  velocity was calculated by averaging over the azimuthal direction at each time step and for the last one, 'equilibrium', $\bar{u}_{\phi}$ is calculated analytically by using the equilibrium equation (\ref{equ:omega}) at each step. 
For the 2D model we have used the equilibrium method as well. 
 }
\label{fig:average}
\end{figure}

To estimate any possible angular momentum transfer caused by the turbulent motions induced by the instability 
we calculate the corresponding Reynolds stress \citep{2003ARA&A..41..555B} 
\begin{equation}
  T_{r \phi} = \frac{\int \rho \delta u_r \delta u_{\phi} dV} 
  {\Delta V} =   \, < \rho \delta u_r \delta u_{\phi}> \,,
\label{eq:Wrp}
\end{equation} 
where $\delta u_r$ and $\delta u_{\phi}$ are defined as the fluctuations of the velocity field from the mean flow
and $\Delta V$ is the volume of the integrated domain.  To calculate a coordinate dependent stress we integrate only over thin slices with a thickness of one cell in the apropriate direction.  While $\delta u_r$ is just the radial velocity, $u_r$, at the point of interest because the initial $u_r$ was zero, $\delta u_{\phi}$ is difficult to calculate, as one has to subtract the mean background rotational velocity.  \citet{2011ARA&A..49..195A} defines it as the difference to the Kepler rotation, while strictly speaking it is the deviation from the unperturbed equilibrium state that is not Keplerian in our case, see eq.~(\ref{equ:omega}). In 3D simulations it is mostly calculated by averaging over the azimuthal direction \citep{2011ApJ...735..122F,2006A&A...457..343F}. But this instability is nearly axisymmetric (see Fig.~\ref{fig:3dmidplane}), so this is not appropriate here and the correct way is to average over time to obtain the steady state velocity. However, this is computationally inconvenient, because this time 
average is not known a priori. In Fig.~\ref{fig:average} we show that the time averaging method leads to the same results as the
equilibrium method using the analytic equation (\ref{equ:omega}), and we use the latter for our subsequent simulations.

To calculate the dimensionless $\alpha$-parameter, $T_{r \phi}$ has to be divided by the pressure. To show the radial and vertical dependence of $\alpha$ it is useful to use different normalizations.  We divide the Reynolds stress in Eq.~(\ref{eq:Wrp}) by the midplane pressure to illustrate the dependence on the meridional (vertical) coordinate, thus making it independent of the number of scale heights of the domain. The stress as a function of the radius, $T_{r \phi}(R)$, is divided by the vertical averaged pressure, making it again independent of the numbers of scale heights. This procedure corresponds to a density weighted height integration \citep{2003ARA&A..41..555B}.

In figure \ref{fig:average} we present the different methods for calculating the Reynolds stress, $T_{r \phi}$, for the simulation of a quarter of a disc with a resolution of 512x128x128 and the same initial conditions as in the 2D case. We can see that indeed the axisymmetric property of the instability leads to wrong results if one only averages over the azimuthal direction. All further results for the isothermal discs are calculated with the equilibrium method. This allows us to approximate the Reynolds stress even in a transient disc and calculate the stress continuously during the whole runtime of the simulation, strongly reducing the amount of data needed to be written to the hard drive, because the Reynolds stress can now be calculated independent of the other time steps.
The computations show in addition that the stresses of the reduced 2D simulations yield stresses comparable to the full 3D case and can be used as a proxy for the full 3D case. 
In Fig. \ref{fig:3dmidplane} we display the vertical velocity in the midplane of the disk for the 3D model.
As shown, the motions are only very weakly non-axisymmetric.

\subsection{Resolution}
\begin{figure}[tb]
\begin{center}
\includegraphics{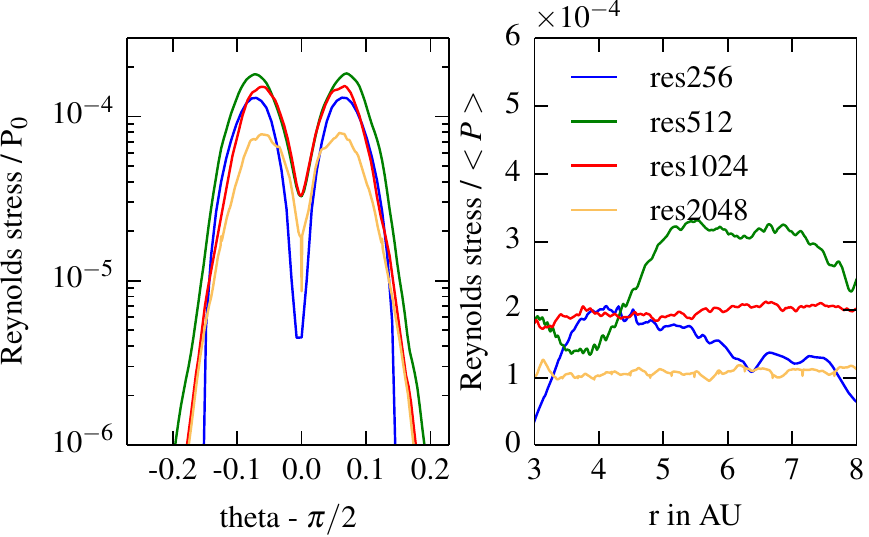}
\end{center}
\caption{Radial and vertical distribution of the Reynolds stress. {\bf Left:} The Reynolds stress divided by the midplane pressure over the vertical direction.
{\bf Right:} Reynolds stress divided by the mean pressure over the radius for different resolutions. Both are averaged over 4001 time steps from orbit 1000 to 5000. The model res2048 corresponds to the results shown in Fig. \ref{fig:growth}. }
\label{fig:alpha_rtheta}
\end{figure}

In this section we take a look at the effect of resolution. We start with a resolution of $256\times64$, where the instability exists but clearly is not resolved and go, by doubling the resolution in several steps, up to a resolution of $2048\times512$, where the computations start to be expensive. In Fig.~\ref{fig:alpha_rtheta} we show on the left the Reynolds stress divided by the midplane pressure as a function of vertical distance. This is then averaged over the radius from $3-8 \AU$. On the right we plot the Reynolds stress divided by the pressure, where both, pressure and stress, have been averaged over the meridional direction. 

From this plot it is not clear if the values for $\alpha$ converge to a specific level for higher resolution. 
But nevertheless it gives a first impression on the strength of turbulent viscosity caused by this instability 
being relatively weak with $\alpha$-values a few times $10^{-4}$, 
which is slightly smaller than the value of $6 \cdot 10^{-4}$ as found by \citet{2013MNRAS.435.2610N}.

\begin{figure}[tb]
\begin{center}
\includegraphics{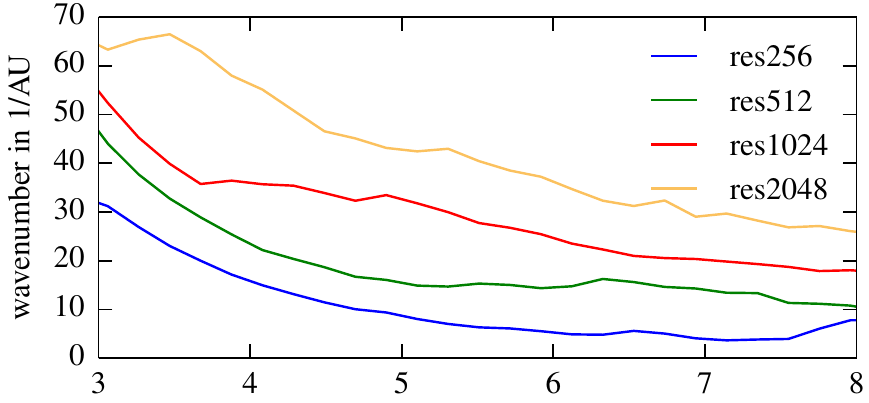}
\includegraphics{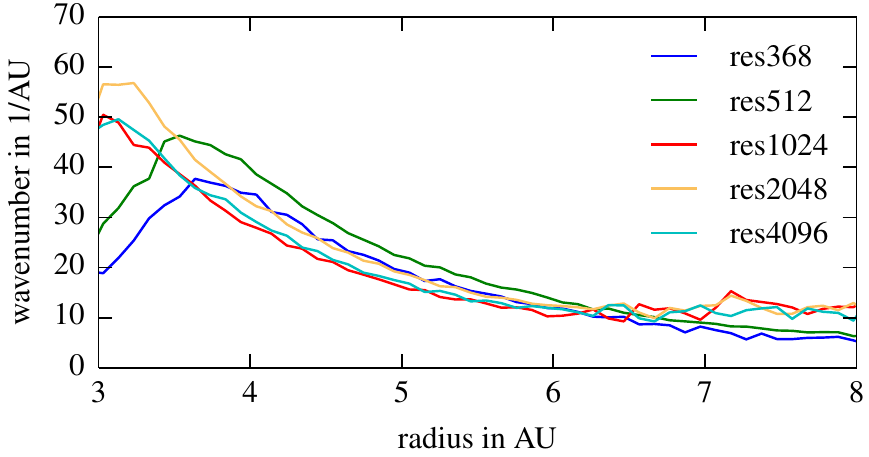}
\end{center}
\caption{The mean wavenumber of the instability over the radius for different numerical resolutions in the saturated phase. 
  Upper panel: inviscid case with $\nu=0$, lower panel: viscous case with $\nu=5 \cdot 10^{-7}$ (dimensionless). 
   }
\label{fig:wave}
\end{figure}

\begin{figure*}[tb]
\begin{center}
\includegraphics{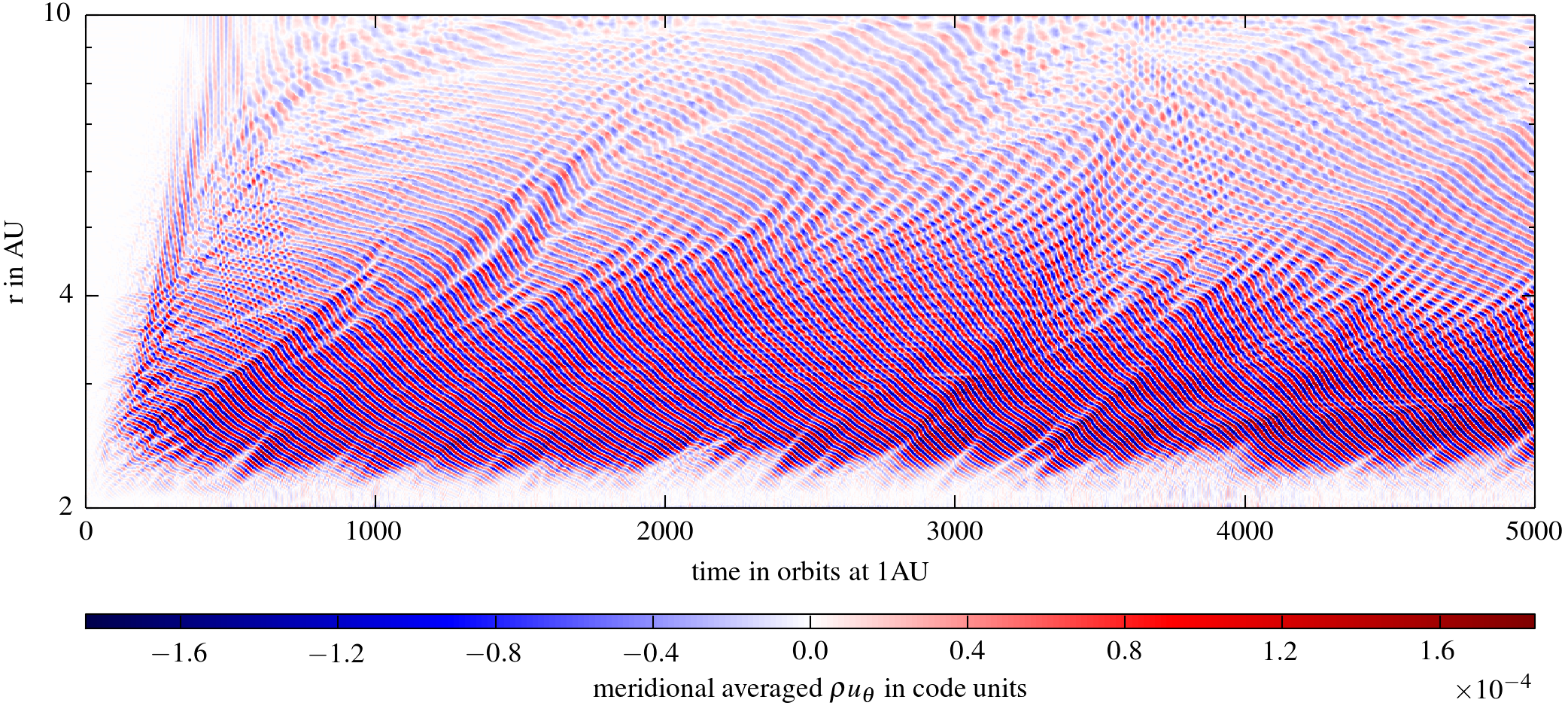}
\end{center}
\caption{Illustration of the large scale time development of the instability. Displayed is the vertically averaged momentum 
in the meridional direction for the inviscid isothermal simulation with a resolution of $1024\times256$ 
(red curve in top panel of Fig.~\ref{fig:wave}).}
\label{fig:overTime}
\end{figure*}

In Fig.~\ref{fig:wave} we display the wavelength of the perturbation as a function of radius for different numerical resolutions, where the wavelength has been estimated by measuring the distance between two sucsessive changes of the sign of the vertically averaged vertical momentum after the instability is saturated (see Fig.~\ref{fig:phase1}, third panel or Fig.~\ref{fig:overTime} along the radius axis, beginning with orbit 1000). 
This does, of course, not reveal the full spectrum, but at this point we are more interested in the characteristic mean wavelength.
Note that the wavelength in the growth phase can be smaller. In all shown resolutions one wavelength is resolved with 15-50 grid cells, while larger radii are better resolved. 
Despite the variation with radius one notices in Fig.~\ref{fig:wave} that the wavelength clearly depends on the numerical resolution.
One possible cause for this is the lack of physical viscosity. Because the (intrinsic) numerical viscosity of the code decreases with increasing resolution, this may explain the missing convergence, in particular since the growth rates depend of the wavenumbers of the disturbances, see eq.~(\ref{eq:sigma0}). 
We repeated the run with an intermediate resolution of $1440 \times 360$ with reduced precision by using a first order instead of a second order spatial interpolation. This clearly increased the wavelength (by about 40\%) indicating that the problem is caused by the numerical viscosity.
\begin{figure}[tb]
\begin{center}
\includegraphics{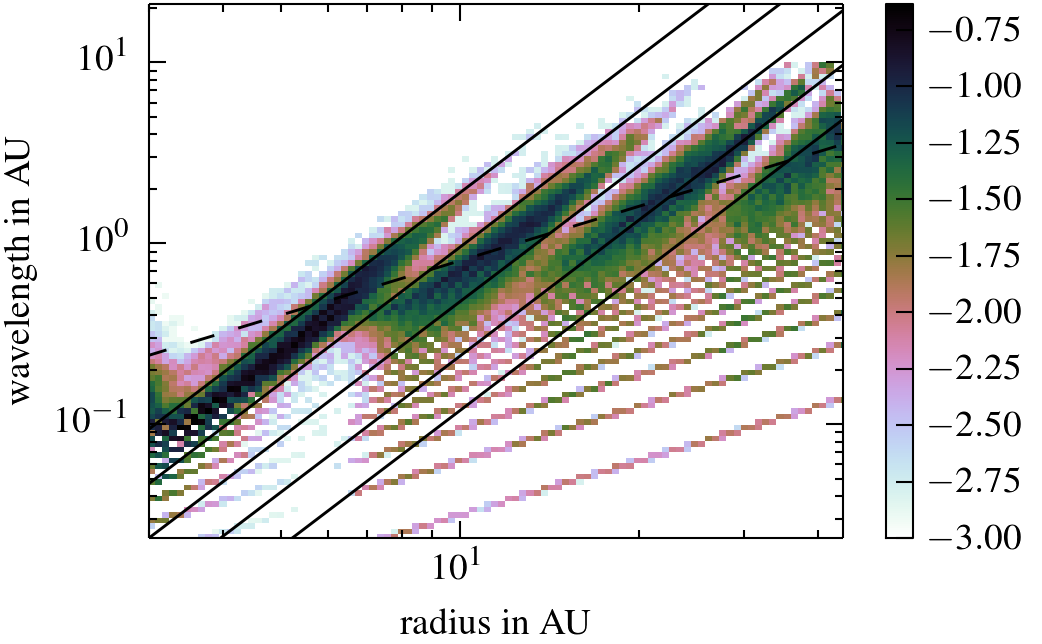}
\end{center}
\caption{Histogram: Color coded is the logarithm of the probability for the occurrence of a wavelength at a radius normalised at each radius by the sum of all wavelengths for the specific radius. The black lines are proportional to the radius to the power of $2.5$ and the lines are a factor of 2 apart from each other. The dashed line has linear slope. One can see that the instability jumps successively between different modes for the wavelength with corresponding jumps in frequency at the same radius. }
\label{fig:large_wave}
\end{figure}

Fig.~\ref{fig:wave} indicates a strong reduction of the wavenumber with radius.
To further explore this dependence of the wavelength on the radius, we performed an additional simulation with an extended radial domain from $2\AU$ to $50\AU$. Again, we estimate the wavelength by measuring the distance between two sign changes in the vertical averaged vertical momentum. This time we show all the wavelengths that were detected by this method in Fig.~\ref{fig:large_wave}, where we display how often a certain wavelength was captured, normalised to the specific radius where it was measured.
An interesting behavior can be observed. While the global radial wavelength does indeed depend linearly on the radius, locally it clearly deviates from this dependence and instead depends on the radius to the power of 2.5. This can also be seen in the simulation with smaller domain, but there it can not be clearly distinguished from the interaction with the boundary.

This supplies us with an explanation for the resolution dependence of the instability. Since the modes can not become arbitrarily small, because of the finite grid, or large, because of the limited vertical scale hight, there will be jumps between different modes. The viscosity and the Kelvin-Helmholtz instability, which can be observed in the simulations with high resolution, are the  candidates for a physical cause for this cut off at small wavelengths.

\begin{figure}[tb]
\begin{center}
\includegraphics{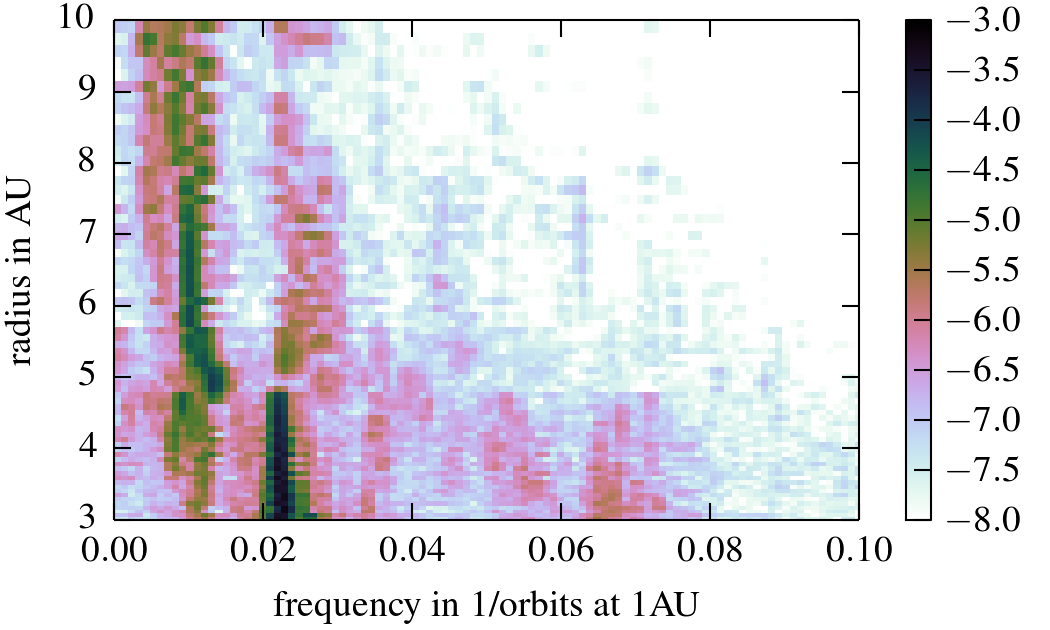}
\end{center}
\caption{
Fourier power spectrum of the temporal evolutionof the instability after saturation (see figure \ref{fig:overTime}). Analysed is the averaged meridional momentum of the simulation without viscosity and resolution $1024x256$.  Color coded is the logarithm of amplitude of the frequency.}
\label{fig:fourier}
\end{figure}

Due to the radius dependence of the wavenumber a spatial Fourier transform is not applicable.
Additionally, as we show bellow, the wavelength in radial direction is not constant in time and also phase jumps can occur.
However, to obtain nevertheless more insight into the dynamics of the system
we display in Fig.~\ref{fig:fourier} the results of a Fourier analysis in time of the vertical momentum of the simulation with resolution of $1024\times256$ (Fig.~\ref{fig:overTime}, along the time axis).
To reduce the problems that phase jumps (see below) pose for the analysis, we step through the data with a Hanning window over 1000 orbits and then average over those 5\% of the resulting spectra that display the highest amplitude. 
We can see a dominant frequency at $0.022 \Omega_K$ at the inner region, this frequency is halved at the outer region beginning at about 5\AU. These jumps in the frequency domain coincide with the jumps in wavenumber. When the wavenumber jumps up, the frequency jumps down, indicating an inverse relationship.
On each branch the frequency is constant, while the wave number varies as $\propto r^{-2.5}$. We can understand this 
relationship starting from eq.~(\ref{eq:sigma0}) from which one obtains for stable inertial oscillations \citep[see eq.~36 in ][]{2013MNRAS.435.2610N}
\begin{equation}
\sigma^2 \sim  -\Omega^2 \frac{k_Z^2}{k_R^2} \,.
\label{eq:oscill}
\end{equation}
The vertical scale is given by the local disk's scale height $H \sim r$ and hence $k_Z \sim r^{-1}$. In the quasi stationary phase, we observed $k_R \sim r^{-2.5}$ (cf. Fig.~\ref{fig:large_wave}), leading to an oscillation frequency independent of the radius, which we also observed (cf. Fig.~\ref{fig:fourier}).

To obtain further insight into the spatio-temporal behaviour of the flow dynamics we display in Fig.~\ref{fig:overTime} the vertically averaged momentum in the vertical direction as a function of space and time.
In this global overview we observe waves that appear to travel slowly from larger to smaller radii.
As noticed already in the Fourier analysis in Fig.~\ref{fig:fourier}, there exists a transition between 4-5\AU\ with
a change in wavelength of the perturbations and occasional phase jumps.
Coupled to this is a change in the typical inward speed of the waves.
They move slower when farther away from the star. As inferred roughly from Fig.~\ref{fig:overTime}, 
the wave speed at r=6\AU\ is about $0.5\AU$ per 250 orbits, while at 4\AU\  it is about $1\AU$. 
However, there is some dependence of this speed on time and space.

Near the outer boundary we see sometimes a region with
standing waves, indicating that the radial domain should not be too small. This
region is mostly only a few wavelengths in size (less then $1\AU$), but can
sometimes also reach a few $\AU$ into the domain. Reflections with the
outer boundary play a role here as well as can be seen in
Fig.~\ref{fig:overTime} for example at $t\approx 500$ or $3600$.
We note that in contrast to our treatment at the inner radial boundary, we did not apply a damping region at the outer boundary.

To check if indeed the viscosity is important for the wavelength we add a small viscosity of $\nu = 5 \cdot 10^{-7}$.  
As expected this leads to a wavenumber that is independent of the resolution, as shown in the bottom panel of Fig.~\ref{fig:wave}.
The wavelength is in the order of $0.2\AU$ at a radius of $4\AU$ after the instability is saturated.

\begin{figure}[tb]
\begin{center}
\includegraphics{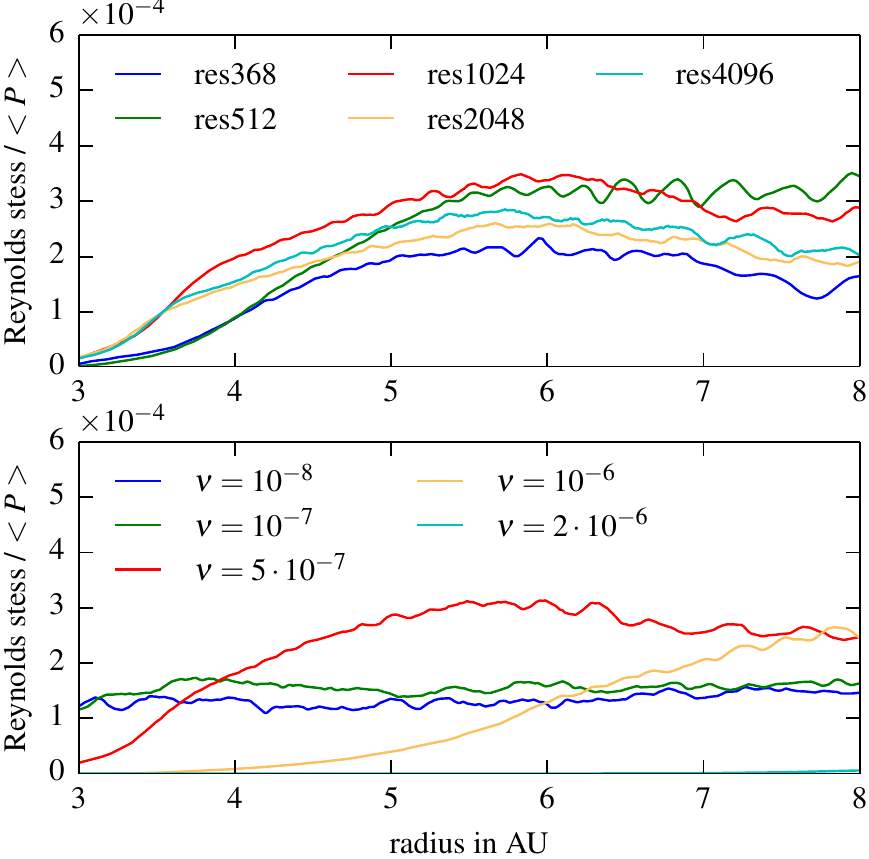}
\end{center}
\caption{The vertically averaged Reynolds stress divided by the vertically averaged pressure in the saturated phase.
Upper panel: For a viscosity of $5 \cdot 10^{-7}$ at different numerical resolutions. 
Lower panel: Fixed resolution of $1440\times360$ and different viscosities. Both
are averaged from orbits 1000 to 3000.}
\label{fig:visc7}
\end{figure}
    
With the wavelength fixed, the Reynolds stress also shows no strong dependence on the resolution as can be seen Fig.~\ref{fig:visc7} top panel. The inner region is strongly suppressed because we also increased the damping from $2-3\AU$.  With that we conclude that a small viscosity is necessary to introduce a physical lengthscale for the smallest unstable wavelength. 
To further explore the role of viscosity we repeat the simulation for different viscosities. This is done with a resolution of $1440\times360$. The growth rate is then calculated by fitting an linear function to the logarithm  of the kinetic energy, which was at each point averaged over 100 grid cells. 
The results in the lower panel of Fig.~\ref{fig:visc7} indicate that the for the two lowest viscosities ($10^{-8}$ and $10^{-7}$) the stresses are given by the numerical viscosity. For the intermediate case ($10^{-7}$) the stresses are larger while for very large values the effect of the increased damping near the inner boundary influences the results.

\section{Discs with radiation transport}
\begin{figure}[tb]
\begin{center}
\includegraphics{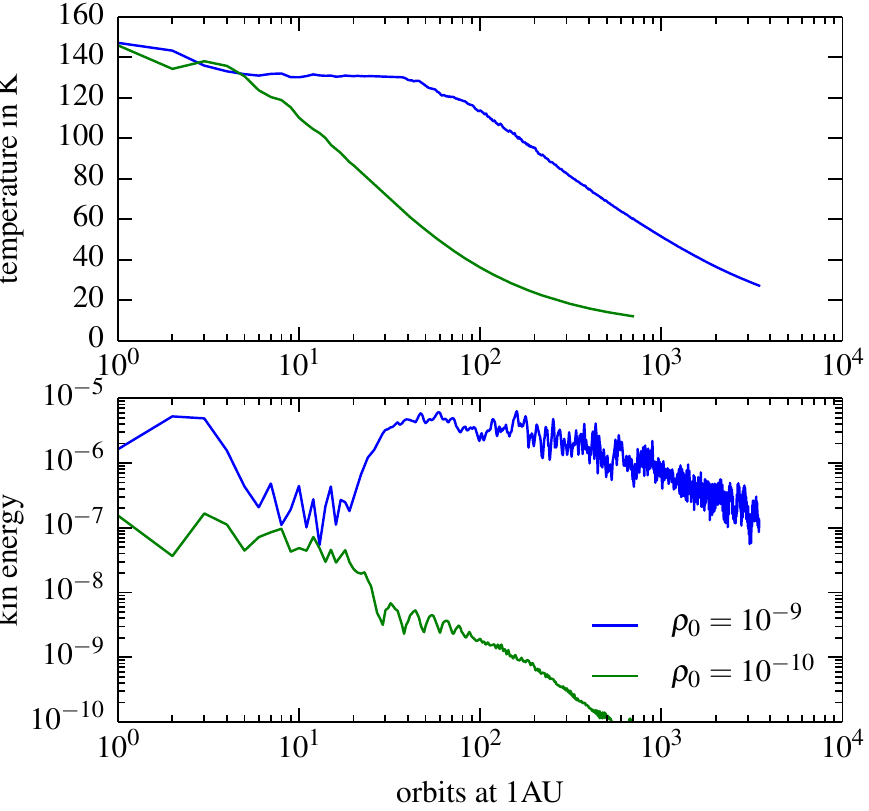}
\end{center}
\caption{Discs with radiation transport for two different densities, $\rho_0$. Upper panel: The midplane temperature at $4\text{-}5\AU$ as a function of time.
Lower panel: Kinetic energy in the meridional flow in the discs.
}
\label{fig:radiation}
\end{figure}
The isothermal discs discussed above do not capture the full physics, most importantly the transport of energy is missing.
In this section we include radiative transport and the heating/cooling interaction of the gas with the radiation.
In the first set of models we start from the isothermal models as described above and switch on the radiation according to eqs.~(\ref{eq:euler3})
and (\ref{eq:rad}), in a second series of models (in Sect.~\ref{sec:irrad}) we include irradiation from the central star.

For the simulations with radiative transport we use a resolution of $1024 \times 256$ and the same spatial extent and initial conditions as in the isothermal case. 
In Fig.~\ref{fig:radiation} we show the midplane temperature averaged from 4-5$\AU$ and the meridional kinetic energy when radiation is included, for two different values of the disc density $\rho_0$.
In both cases the kinetic energy has initially larger amplitudes than in the previous isothermal simulations because now the
disc is no longer in hydrostatic equilibrium initially, and small motions in the meridional plane set in (lower panel in Fig.~\ref{fig:radiation}).
For the same disc density as before, $\rho_0 = 10^{-10}$, the disc cools off quickly as soon as the instability begins to be active, at around $t=10$.
The reason lies in the efficient radiative cooling in this case, in particular near the surface layers where the optical depth is small
and the instability most active. Hence, any turbulent heating will be radiated away quickly.

We repeated the simulation with a higher density, $\rho_0 = 10^{-9}$ at $1\AU$, to increase the optical thickness.
Now the disc does not cool so efficiently such that the instability begins to set in between $t=10$ and $t=20$ orbits, very similar to
the isothermal models. But then radiative cooling eventually leads again to a cooling of the disc and the instability
dies out.
From these results it is clear that the instability does not produce enough heat and cannot survive without an external source of heat, for the typical
opacities and densities expected in protoplanetary discs. This potential problem was pointed out already by \citet{2013MNRAS.435.2610N}.

\section{Irradiated Disc}
\label{sec:irrad}
Here, we extend our models and include irradiation from the central star as an external heat source. 
Of course there are also other sources possible, for example, the inner region of the disc where the MRI is still active could be important.

\subsection{Method of irradiation}
We use a simple model for the external heating and consider vertical irradiation from above and below the disc, where the energy flux, $\vec{F}^{irr}$, depends on radius.
This procedure evades the problem of finding a self-consistent solution for the flaring of the disc,
as done for an irradiated and internally heated disc by \citet{2013A&A...549A.124B}.

To obtain a first approximation for the flux in the meridional direction we
assume that the angle of incidence of the flux is approximately
${R_{\odot}}/{r}$, where $R_{\odot}$ is the star's radius. 
This applies to an infinitely flat disc as well as to the upper and lower surfaces our computational
grid in spherical polar coordinates because all three represent planes that cross the center of the central star.
We obtain for the meridional component of the flux 
\begin{equation}
   F^{irr}_{\theta}  = F_{r} \,  \frac{R_{\odot}}{R} \,,
\label{equ irrad}
\end{equation}
where $F_{r} = F_{\odot} (R_{\odot}/r)^2$ is the radial flux from the star at a distance $r$.
Applying this impinging vertical irradiation to the disc leads to a radial temperature profile exponent in the disc of $q=-0.55$,
in good agreement with the models of \citet{1997ApJ...490..368C}. 
Our procedure does not allow for self-shadowing effects \citep{2013A&A...549A.124B} but should give a physically realistic estimate of the
expected temperatures in the disc.

For the irradiation opacity we use here, to simplify the calculations and obtain a first order estimate of the effect, the same Rosseland opacity of \citet{Bell1994}. Hence, in the simulations we use presently the same opacity for the irradiation, Rosseland and Planck opacity \citep{2013A&A...549A.124B}. Numerically, we perform a ray-tracing
method to calculate the energy deposited in each cell of the computational grid \citep{2013A&A...559A..80K}. 

\subsection{Growth rate}
To measure the growth rates of the instability for discs with radiation transport and
irradiation we ran models for with zero viscosity and a higher resolution case with $\nu=10^{-7}$.  
To be able to compare the growth rates with the previous isothermal cases, we
performed additional isothermal simulations using the temperature profile from the simulations with radiation transport and irradiation.

The growth rates for the instability in combination with radiation transport are difficult to capture, because the simulation can not be started in hydrostatic equilibrium because the equilibrium vertical profile is unknown. We use strong damping for the first 10 orbits to remove the disturbance caused by the transition to the new density and temperature profile.

\begin{figure}[tb]
\begin{center}
\includegraphics{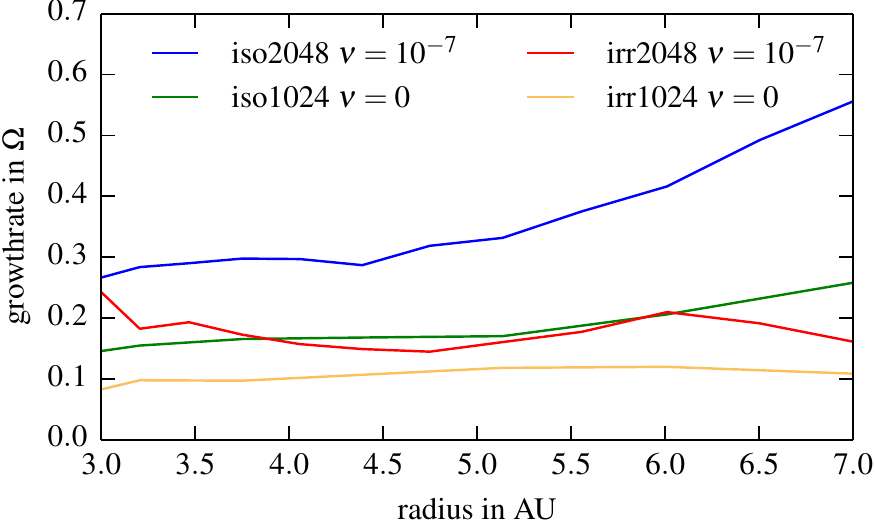}
\end{center}
\caption{Growth rates for models with radiation transport and irradiation (irr) and isothermal models (iso) that have the same mean
temperature, for comparison. For each radius the kinetic energy was smoothed over a range of 10\% of its radius before fitting it to an exponential growth.}
\label{fig:sigma_irr}
\end{figure}
\begin{figure}[tb]
\begin{center}
\includegraphics{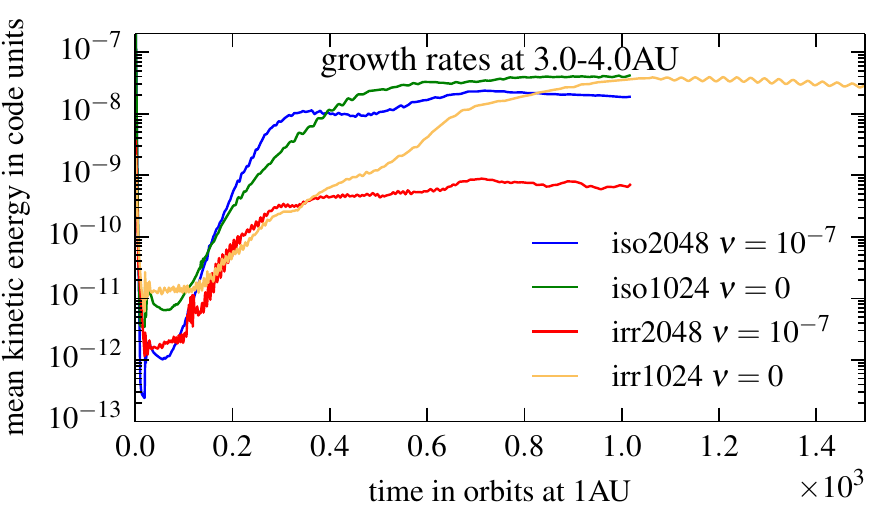}
\includegraphics{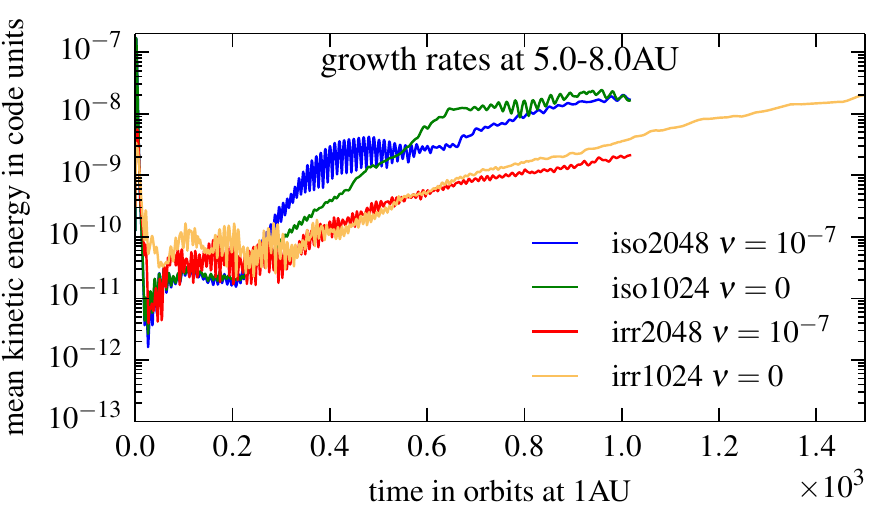}
\end{center}
\caption{Growth of the kinetic energy in the 2D-plane with radiation transport and isothermal for comparison. }
\label{fig:growth_irr}
\end{figure}

The results are shown in Fig.~\ref{fig:sigma_irr}. We note that this time the growth rates should depend on radius, because the growth depends on $H/R$ which is not constant in the radiative cases. From Fig.~\ref{fig:sigma_irr} it is clear that the growth rates for the isothermal models are now lower than in the cases presented above, because the temperature is first lower and second the radial profile is flatter as before, and both are important for growth.
For the irradiated models the growth is again lower, with $0.1-0.2$ per local orbit around half the value for the isothermal case.
In Fig.~\ref{fig:growth_irr} we display the evolution of the kinetic energy for the irradiated and corresponding isothermal model.
For the inviscid case the final saturation level agree very well with each other while for the viscous disc with $\nu = 10^{-7}$ the instability
is weaker in the inner regions of the disc (see below).

\subsection{Quasistationary phase}

\begin{figure}[t]
\begin{center}
\includegraphics{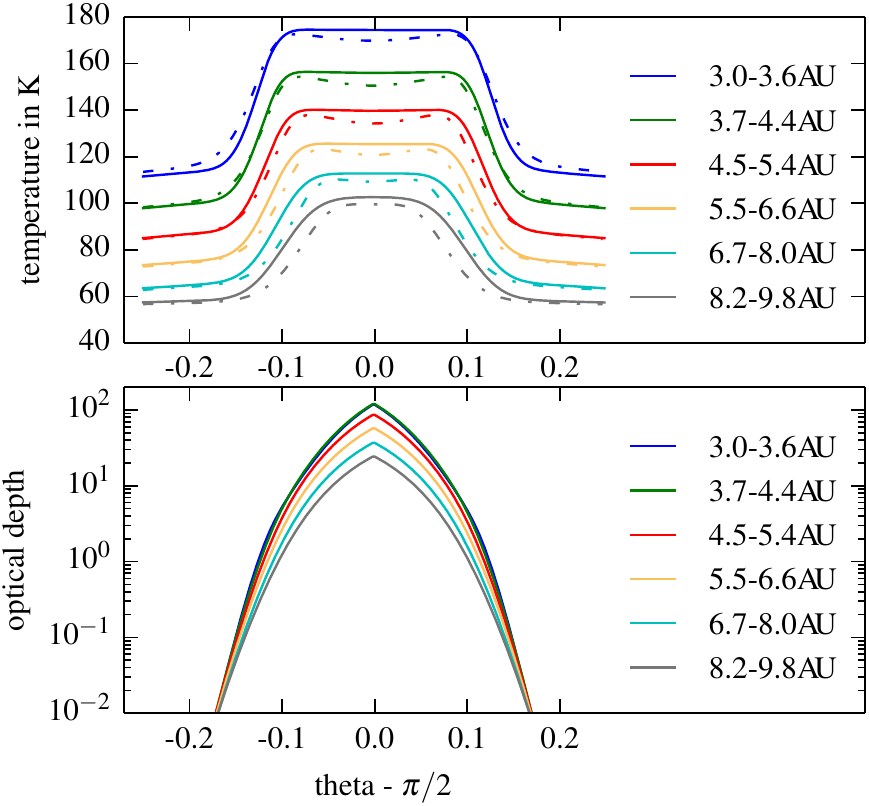}
\end{center}
\caption{Upper Panel: The temperature profile for an irradiated disc in the saturated phase
without viscosity and a resolution of $1024 \times 256$. The dotted line is a run without hydrodynamics, only solving for the radiation energy. Lower Panel: The vertically integrated optical depth.}
\label{fig:irrad_temperature}
\end{figure}

In the top panel of Fig.~\ref{fig:irrad_temperature} we display the vertical temperature distribution for the
saturated state at different radii in the disc for the model without viscosity at a resolution of $1024\times 256$. 
The other models look very similar.  In the bulk part of the disc the profile is quite flat with a slight drop towards the midplane.
A lower central temperature is to be expected for irradiated dics, detailed radiative transfer models indicate an even larger temperature drop towards the midplane \citep{2002A&A...389..464D}.  In the upper layers the temperature falls off because the disc is optically thin and the energy can freely leave the system. This drop of the temperatures towards the surface despite the irradiation is a result of the
identical irradiation and Rosseland opacity. If more radiation is allowed to be absorbed in the disk by increasing the irradiation opacity
then one can obtain hotter surface layers. For a ten times larger value we find a hot corona similar to \citet{2013A&A...560A..43F} and a cooler midplane. First results seem to indicate a reduction in the Reynolds stress in this case, probably due to the lower temperature in the bulk of the disc. At this point we leave the details to subsequent
studies.
The dotted line in Fig.~ \ref{fig:irrad_temperature} shows the profile for a simulation where we only solve for the radiation energy and disable the hydrodynamic solver. We can infer from this that the flat profile is a result of the combination of turbulent heating and vertical motion.
A test simulation with a passive tracer added in the midplane of the disk in the saturated state showed indeed rapid spreading over the whole vertical extent of the disk.

\begin{figure}[tb]
\begin{center}
\includegraphics{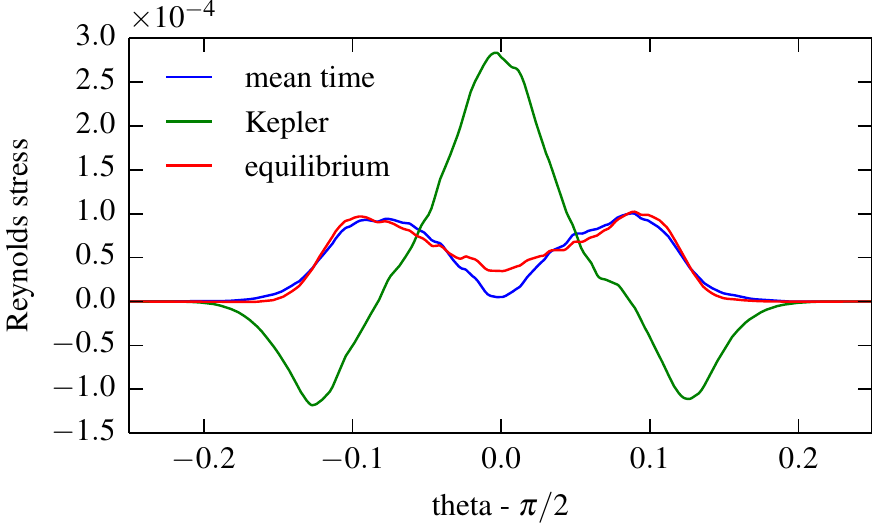}
\end{center}
\caption{Irradiated run: The Reynolds stress was averaged over 41 time steps,
from orbit 1000 to 5000 each step 100 orbits apart, calculated with different
averaging methods. For 'mean time' the mean $u_{\phi}$ was calculated through
averaging over 40 time steps. For 'Kepler' the velocity was calculated by
subtracting the Kepler velocity and for the last one $u_{\phi}$ is calculated
analytically by using the equilibrium equation (\ref{equ:omega}). Spatial
averages are taken from $4\AU$ to $10\AU$.}
\label{fig:average_irr}
\end{figure}

The vertically integrated optical depth is shown in the lower panel, starting from very small values at the disc surfaces it reaches
30-100 at the different radii.
The nearly constant vertical temperature within the disc motivates us to use the equilibrium azimuthal velocity for the corresponding
isothermal model of the steady state to calculate the Reynolds stress. In Fig.~\ref{fig:average_irr} we can see that it is still a good approximation.
Note that this time the comparison is done with a 2D-simulation. Also shown is the Reynolds stress calculated with the Kepler velocity instead of the equilibrium velocity.

\begin{figure}[t]
\begin{center}
\includegraphics{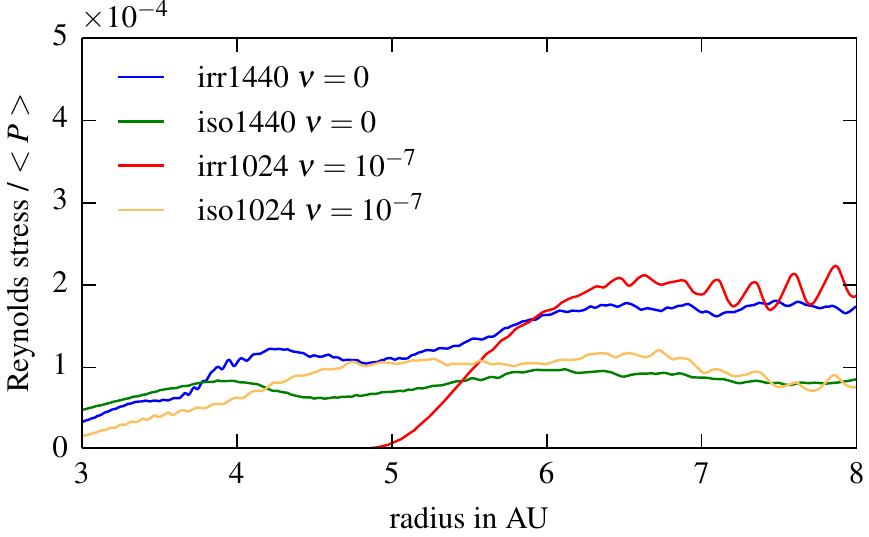}
\end{center}
\caption{Comparison of the Reynolds stress divided by the mean pressure. 'irr' stands for the irradiated disc, 'iso' for the isothermal disc with analogous initial conditions. Averaged over 2000 orbits, beginning with orbit 2000.}
\label{fig:irrad_over_r}
\end{figure}

While the growth rates are weaker than in the isothermal case, the kinetic energy in the meridional plane for stable saturated phase in Fig.~\ref{fig:irrad_over_r} reaches the same level with radiation transport. The values for $\alpha$ are again between $0.5 \cdot 10^{-4}$ and $2 \cdot 10^{-4}$, depending on the wavelength and thus viscosity, but independent on radiation transport. 

Different is of course the strength of the instability measured in terms of the value of the viscosity under which it still survives.
Here, in the irradiated case, even a low viscosity of $10^{-7}$ does suppress the instability in the inner regions of the disc.
This is not only a result of the radiation transport, but also of the flat temperature profile. 
The details will depend on the source of the heating and the opacity, but nevertheless the stability will be weaker than in the purely isothermal case.

\subsection{Discussion}
\begin{figure}[tb]
\begin{center}
\includegraphics{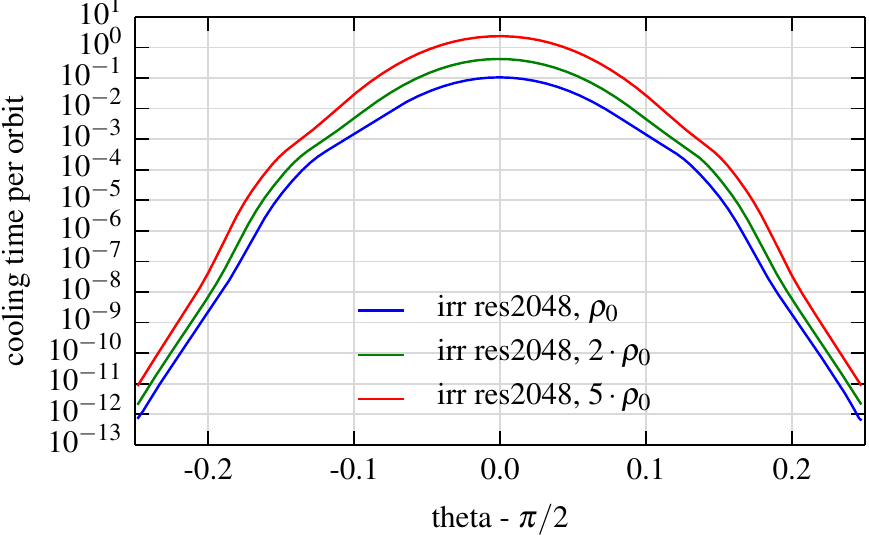}
\end{center}
\caption{The radiative diffusion time per Orbit at $4 \AU$ for a lengthscale of $0.1\AU$}
\label{fig:cool}
\end{figure}

As we have shown in the previous sections for an irradiated disc there exists the possiblity of generating an effective turbulence through the
vertical shear instability. As pointed out in \citet{2013MNRAS.435.2610N} the instability can only be sustained if the diffusion
(local relaxation) time is a fraction of the local orbital period. To investigate how this condition is fulfilled in our simulations 
we analyse for the equilibrium irradiated disc models the timescale for radiative diffusion. 
In units of the local orbital period this is given by
\begin{equation}
 t_\textrm{diff}  =  \Delta x^2 \frac{ c_P \rho^2 \kappa_R}{4 \lambda a c T^3} \cdot \frac{\Omega}{2 \pi}
\label{eq:tdiff}
\end{equation}
where $\Delta x$ is the characteristic wavelength of the perturbation. In our case the radial diffusion is relevant \citep{2013MNRAS.435.2610N}
and we choose here $\Delta x = 0.05 r$, which is a typical radial wavelength at $r=3\AU$. 
Using eq.~(\ref{eq:tdiff}) and the results from the simulation we calculate for the optical thin region a very small cooling time per
orbit of $t_\textrm{diff} = 10^{-10}$ as expected. For the optical thick region we obtain $t_\textrm{diff} = 0.11$ for our standard density,
which is indeed a small fraction of the orbital period as required for the instability to operate, see Fig.~\ref{fig:cool}.
The cooling time in the vertical direction is longer, about a few orbital periods
as implied by the vertical optical depth (see Fig.~\ref{fig:irrad_temperature}) but this will keep the disc nearly isothermal,
again as required for instability.
 
\begin{figure}[tb]
\begin{center}
\includegraphics{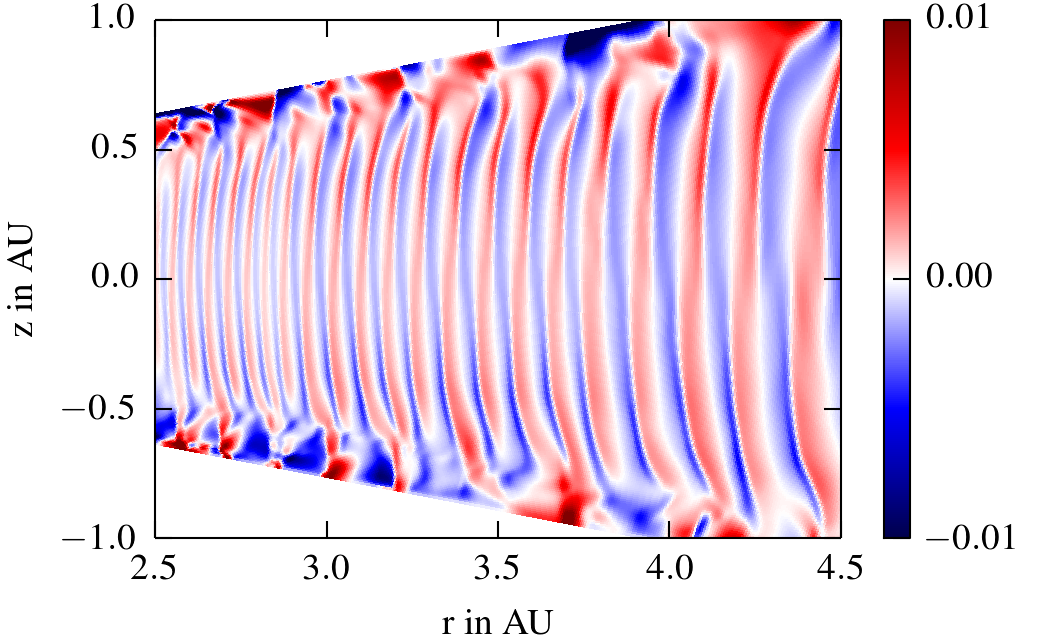}
\includegraphics{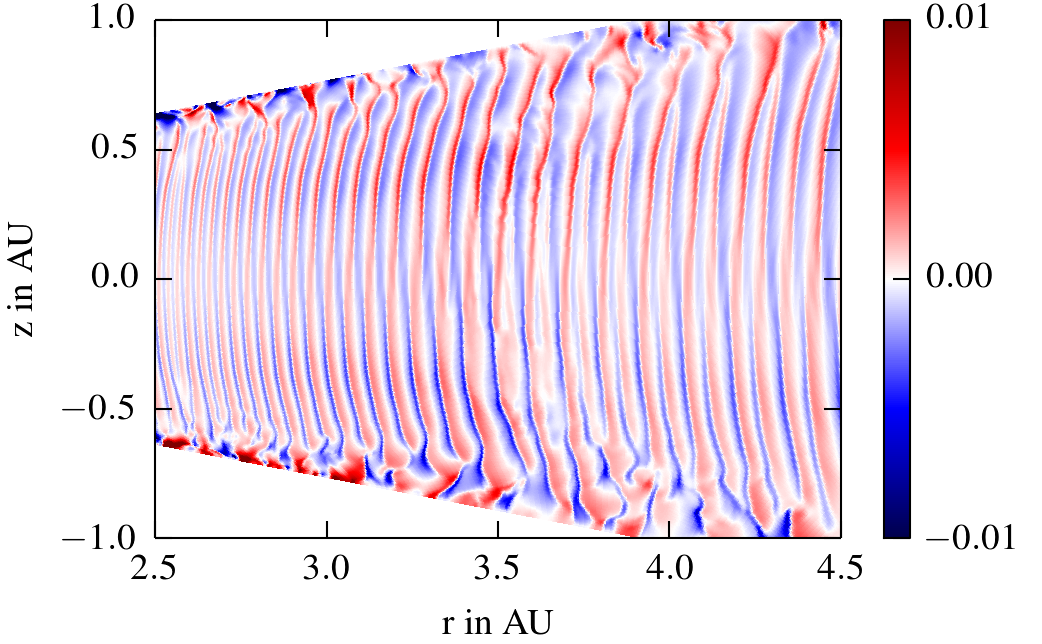}
\end{center}
\caption{Velocity in the meridional direction, $u_\theta$, in units of local Kepler velocity for an irradiated run without viscosity at resolution $1024\times256$ and below with resolution $2048\times512$. Compare with Fig. \ref{fig:phase1} which has a spatial resolution of $2048\times512$.}
\label{fig:phase_irr}
\end{figure}

In Fig.~\ref{fig:phase_irr} we illustrate that the instability still resembles closely the locally isothermal case except that the small
scale perturbations are missing, even in the optical thin region, where we have very small cooling times. 
For comparison, \citet{2013MNRAS.435.2610N} found that the instability was completely suppressed with relaxation times of $t_\textrm{relax} = 0.1$,
which is the timescale for the flow to relax to the initial isothermal profile.
We take this as an indication that radiative diffusion plus irradiation behaves physically different from a simple model of temperature relaxation
as used in \citet{2013MNRAS.435.2610N}. 

As seen in Fig.~\ref{fig:irrad_temperature} an increase in the density leads to higher optical depths 
and longer diffusion times, and consequently to a weaker instability. 
While doubling the density in a simulation with resolution $2048\times512$ has no clear influence on the kinetic energy and the cooling times 
in the optical thin regions, the Reynolds stress was clearly weaker by a factor of around 1.5 in the simulation with doubled density 
(the model in the middle of Fig.~\ref{fig:cool}). In addition the wavelength  of the perturbations is decreased.

A further increase of the density leads also to a strong decrease in the kinetic energy, with again a smaller wavelength. 
This raises the question whether the simulation with resolution of $2048\times512$ is sufficiently resolved. 
These results indicate that in very massive discs with long diffusion times (vertical and radial) the disc will behave more adiabatically, and the
instability will be quenched.
The minimum solar mass nebula corresponds at 5 AU approximately to our model with 2 $\rho_0$ and the instability might just be operative.

\section{Summary and conclusions}
We have studied the vertical shear instability (VSI) as a source of turbulence in protoplanetary discs.
For that purpose we have performed numerical simulations solving the equations of hydrodynamics for a grid
section in spherical polar coordinates. To study the global behaviour of the instability we have used an large
radial extension of the grid ranging from 2 to 10 AU. 

In a first set of simulations we show that the instability occurs for locally isothermal discs where the
radial temperature gradient is a given function of radius. Our results on the growth rates for the instability 
are in good agreement with the theoretical estimates by \citet{1998MNRAS.294..399U, 2003A&A...404..397U}, and we
find two basic growth regimes for the asymmetric and antisymmetric modes as seen by \citet{2013MNRAS.435.2610N}.
After 20 to 30 local orbits the instability saturates and is dominated by the vertical motions, which cover the
whole vertical extent of the disc. 

Interestingly, we find that the local radial wavelength of the perturbations scales approximately with $\lambda \propto r^{2.5}$ in the saturated state with a constant frequency.
However, on a global scale several jumps occur where the wavelengths are halved, such that the global scaling follows $\bar{\lambda} \propto r$ with $\bar{\lambda}/r = 0.03$. We suspect that the instability has the tendency to generate global modes that show the observed
wavelengths behaviour according to eq.~(\ref{eq:oscill}). Due to the radial stratification of the disc jumps have to occur
at some locations.

The waves approximately keep their shape and travel slowly inwards.
The two- and three dimensional simulations yield essentially the same results concerning the growth rates and saturation levels 
of the instability because of its axisymmetric property.
The motions give rise to a finite level of turbulence and we calculate the associated efficiency, measured in terms of $\alpha$.
We first show that, caused be the two-dimensionality, $\alpha$ can be measured directly from the two-dimensional simulations using
the proper equilibrium state of the disc.
We find that the angular momentum associated with the turbulence is positive and reaches $\alpha$-values of a few $10^{-4}$.
For the isothermal simulations we find that upon higher numerical resolution $\alpha$ becomes smaller but viscous simulations
indicate a saturation at a level of about $\alpha = 10^{-4}$ even for very small underlying viscosities that are equivalent to $\alpha < 10^{-6}$.

Adding radiative transport leads to a cooling from the disc surfaces and the instability dies out subsequently. We then
constructed models where the disc is irradiated from above and below which leads to a nearly constant vertical temperature
profile within the disc. This leads again to a turbulent saturated state with a similar transport efficiency as the purely
isothermal simulations, possibly slightly higher (see Fig.~\ref{fig:irrad_over_r}).

In summary, our simulations indicate that the VSI can indeed generate turbulence in discs albeit at a relatively low level of about 
few times $10^{-4}$. This implies that even in (magnetically) dead zones the effective viscosity in discs will never fall below
this level. Our results indicate that in fully 3D simulations the transport may be marginally larger, but further simulations
will have to be performed to clarify this point.

\begin{acknowledgements}
Moritz Stoll received financial support from the Landesgraduiertenf\"orderung
of the state of Baden-W\"urttemberg.
Wilhelm Kley acknowledges the support of the German Research
Foundation (DFG) through grant KL 650/8-2 within the Collaborative Research
Group FOR 759: The formation of Planets: The Critical First Growth Phase.
Some simulations were performed on the bwGRiD cluster in T\"ubingen, which is
funded by the Ministry for Education and Research of Germany and the Ministry
for Science, Research and Arts of the state Baden-W\"urttemberg, and the cluster
of the Forschergruppe FOR 759 “The Formation of Planets: The Critical First
Growth Phase” funded by the DFG.
\end{acknowledgements}

\bibliography{library,refs}{}

\begin{thebibliography}{28}
\expandafter\ifx\csname natexlab\endcsname\relax\def\natexlab#1{#1}\fi

\bibitem[{{Arlt} \& {Urpin}(2004)}]{2004A&A...426..755A}
{Arlt}, R. \& {Urpin}, V. 2004, \aap, 426, 755

\bibitem[{{Armitage}(2011)}]{2011ARA&A..49..195A}
{Armitage}, P.~J. 2011, \araa, 49, 195

\bibitem[{{Balbus}(2003)}]{2003ARA&A..41..555B}
{Balbus}, S.~A. 2003, \araa, 41, 555

\bibitem[{Bell \& Lin(1994)}]{Bell1994}
Bell, K.~R. \& Lin, D. N.~C. 1994, ApJ, 427, 987

\bibitem[{{Bitsch} {et~al.}(2013){Bitsch}, {Crida}, {Morbidelli}, {Kley}, \&
  {Dobbs-Dixon}}]{2013A&A...549A.124B}
{Bitsch}, B., {Crida}, A., {Morbidelli}, A., {Kley}, W., \& {Dobbs-Dixon}, I.
  2013, \aap, 549, A124

\bibitem[{{Chiang} \& {Goldreich}(1997)}]{1997ApJ...490..368C}
{Chiang}, E.~I. \& {Goldreich}, P. 1997, \apj, 490, 368

\bibitem[{{de Val-Borro} {et~al.}(2006){de Val-Borro}, {Edgar}, {Artymowicz},
  {Ciecielag}, {Cresswell}, {D'Angelo}, {Delgado-Donate}, {Dirksen}, {Fromang},
  {Gawryszczak}, {Klahr}, {Kley}, {Lyra}, {Masset}, {Mellema}, {Nelson},
  {Paardekooper}, {Peplinski}, {Pierens}, {Plewa}, {Rice}, {Sch{\"a}fer}, \&
  {Speith}}]{2006MNRAS.370..529D}
{de Val-Borro}, M., {Edgar}, R.~G., {Artymowicz}, P., {et~al.} 2006, \mnras,
  370, 529

\bibitem[{{Dullemond} {et~al.}(2002){Dullemond}, {van Zadelhoff}, \&
  {Natta}}]{2002A&A...389..464D}
{Dullemond}, C.~P., {van Zadelhoff}, G.~J., \& {Natta}, A. 2002, \aap, 389, 464

\bibitem[{{Flaig} {et~al.}(2012){Flaig}, {Ruoff}, {Kley}, \&
  {Kissmann}}]{2012MNRAS.420.2419F}
{Flaig}, M., {Ruoff}, P., {Kley}, W., \& {Kissmann}, R. 2012, \mnras, 420, 2419

\bibitem[{{Flock} {et~al.}(2011){Flock}, {Dzyurkevich}, {Klahr}, {Turner}, \&
  {Henning}}]{2011ApJ...735..122F}
{Flock}, M., {Dzyurkevich}, N., {Klahr}, H., {Turner}, N.~J., \& {Henning}, T.
  2011, \apj, 735, 122

\bibitem[{{Flock} {et~al.}(2013){Flock}, {Fromang}, {Gonz{\'a}lez}, \&
  {Commer{\c c}on}}]{2013A&A...560A..43F}
{Flock}, M., {Fromang}, S., {Gonz{\'a}lez}, M., \& {Commer{\c c}on}, B. 2013,
  \aap, 560, A43

\bibitem[{{Fricke}(1968)}]{1968ZA.....68..317F}
{Fricke}, K. 1968, \zap, 68, 317

\bibitem[{{Fromang} \& {Nelson}(2006)}]{2006A&A...457..343F}
{Fromang}, S. \& {Nelson}, R.~P. 2006, \aap, 457, 343

\bibitem[{{Goldreich} \& {Schubert}(1967)}]{1967ApJ...150..571G}
{Goldreich}, P. \& {Schubert}, G. 1967, \apj, 150, 571

\bibitem[{{Klahr} \& {Bodenheimer}(2003)}]{2003ApJ...582..869K}
{Klahr}, H.~H. \& {Bodenheimer}, P. 2003, \apj, 582, 869

\bibitem[{{Kley} {et~al.}(1993){Kley}, {Papaloizou}, \&
  {Lin}}]{1993ApJ...416..679K}
{Kley}, W., {Papaloizou}, J.~C.~B., \& {Lin}, D.~N.~C. 1993, \apj, 416, 679

\bibitem[{{Kolb} {et~al.}(2013){Kolb}, {Stute}, {Kley}, \&
  {Mignone}}]{2013A&A...559A..80K}
{Kolb}, S.~M., {Stute}, M., {Kley}, W., \& {Mignone}, A. 2013, \aap, 559, A80

\bibitem[{{Levermore} \& {Pomraning}(1981)}]{1981ApJ...248..321L}
{Levermore}, C.~D. \& {Pomraning}, G.~C. 1981, \apj, 248, 321

\bibitem[{{Lin} \& {Pringle}(1987)}]{1987MNRAS.225..607L}
{Lin}, D.~N.~C. \& {Pringle}, J.~E. 1987, \mnras, 225, 607

\bibitem[{{Mignone} {et~al.}(2007){Mignone}, {Bodo}, {Massaglia}, {Matsakos},
  {Tesileanu}, {Zanni}, \& {Ferrari}}]{2007ApJS..170..228M}
{Mignone}, A., {Bodo}, G., {Massaglia}, S., {et~al.} 2007, \apjs, 170, 228

\bibitem[{{Minerbo}(1978)}]{1978JQSRT..20..541M}
{Minerbo}, G.~N. 1978, \jqsrt, 20, 541

\bibitem[{{Nelson} {et~al.}(2013){Nelson}, {Gressel}, \&
  {Umurhan}}]{2013MNRAS.435.2610N}
{Nelson}, R.~P., {Gressel}, O., \& {Umurhan}, O.~M. 2013, \mnras, 435, 2610

\bibitem[{{Pringle}(1981)}]{1981ARA&A..19..137P}
{Pringle}, J.~E. 1981, \araa, 19, 137

\bibitem[{{Ruden} {et~al.}(1988){Ruden}, {Papaloizou}, \&
  {Lin}}]{1988ApJ...329..739R}
{Ruden}, S.~P., {Papaloizou}, J.~C.~B., \& {Lin}, D.~N.~C. 1988, \apj, 329, 739

\bibitem[{Shakura \& Sunyaev(1973)}]{Shakura1973}
Shakura, N. \& Sunyaev, R. 1973, A\&A, 24, 337

\bibitem[{{Turner} {et~al.}(2014){Turner}, {Fromang}, {Gammie}, {Klahr},
  {Lesur}, {Wardle}, \& {Bai}}]{2014arXiv1401.7306T}
{Turner}, N.~J., {Fromang}, S., {Gammie}, C., {et~al.} 2014, ArXiv e-prints

\bibitem[{{Urpin}(2003)}]{2003A&A...404..397U}
{Urpin}, V. 2003, \aap, 404, 397

\bibitem[{{Urpin} \& {Brandenburg}(1998)}]{1998MNRAS.294..399U}
{Urpin}, V. \& {Brandenburg}, A. 1998, \mnras, 294, 399

\end{thebibliography}
\bibliographystyle{aa}
\end{document}